\documentstyle[aps,eqsecnum,epsf,preprint,tighten]{revtex}
\newcommand{\bk}{{\bf k}}
\newcommand{\bx}{{\bf x}}
\begin{document}
\widetext
\title{Non-zero temperature transport near quantum critical points}
\author{Kedar Damle and Subir Sachdev}
\address{
Department of Physics, P.O. Box 208120,
Yale University, New Haven, CT 06520-8120}
\date{April 26, 1997}
\preprint{cond-mat/9705206}
\maketitle
\begin{abstract}
We describe the nature of charge transport at non-zero temperatures ($T$)
above the two-dimensional ($d$) superfluid-insulator quantum critical point.
We argue that the transport is characterized by inelastic collisions
among thermally excited carriers at a rate of order $k_B T/\hbar$. 
This implies that the 
transport at frequencies $\omega \ll k_B T/\hbar$ is in the hydrodynamic,
collision-dominated (or `incoherent')
regime, while $\omega \gg k_B T/\hbar$ is the collisionless (or `phase-coherent')
regime. The conductivity is argued to be $e^2 / h$ times a non-trivial universal
scaling function of $\hbar \omega / k_B T$, and not independent of
$\hbar \omega/k_B T$, as has been previously claimed, or implicitly assumed. 
The experimentally
measured d.c. conductivity is the hydrodynamic
$\hbar \omega/k_B T \rightarrow 0$
limit of this function, and is a universal number times $e^2 / h$, even though
the transport is incoherent.
Previous work determined the conductivity by incorrectly
assuming it was also equal to the collisionless 
$\hbar \omega/k_B T \rightarrow \infty$ limit of the scaling function,
which actually describes phase-coherent transport with a conductivity
given by a {\em different\/} universal number times $e^2 / h$.
We provide the first computation of the universal d.c. conductivity
in a disorder-free boson model, along with explicit crossover functions,
using a quantum Boltzmann equation and an expansion in $\epsilon=3-d$. 
The case of spin
transport near quantum critical points in antiferromagnets is also discussed. Similar
ideas should apply to the transitions in quantum Hall systems and to metal-insulator
transitions.  We suggest experimental
tests of our picture and speculate
on a new route to self-duality at two-dimensional quantum critical points.
\end{abstract}
\pacs{PACS numbers:}
\widetext
\newpage
\section{Introduction}
\label{intro}
The charge transport properties of systems near a zero temperature quantum phase
transition have been the subject of a large number of experimental studies in the
last few years. Systems in two spatial dimensions have been of special interest,
and important examples are the superfluid-insulator transition in disordered 
thin films~\cite{hebard,lg,yk,cv,ajr} and Josephson junction arrays~\cite{jj}, the transitions 
from quantized Hall plateaus~\cite{shahar,sankar}, 
a quantum critical point in the doped cuprate compounds~\cite{gb},
and a recent
quantum transition in $Si$ MOSFETs~\cite{sarachik}. 
In three dimensions, the metal-insulator transition in doped semiconductors
has seen many years of study~\cite{bk}, and evidence for scaling collapse of data near
the quantum critical point has finally begun to emerge~\cite{lohneysen}.
 We shall present the results
of this paper using the language of the superfluid-insulator transition in thin films
in zero external magnetic field, but we believe our ideas are much more general
and also have significant implications for the other systems noted above. 
We shall also discuss the extension of our results to the case of {\em spin\/}
transport near quantum critical points in two-dimensional antiferromagnets~\cite{csy}.

A particular focus of the experiments on the superfluid-insulator transition
has been the theoretical prediction, made in the seminal work of 
Fisher~{\em et al.}~\cite{fgg}
and Cha~{\em et al.}~\cite{cha}, that the conductivity at the critical point
in two dimensions is non-zero and equals the quantum unit of conductance ($e^2 / h$)
times a universal number.
However there remains an important
dichotomy between experimental and theoretical studies of the superfluid-insulator
transition in disordered thin films, which is crucial to our discussion.
All experiments are
performed at a low, but non-zero, temperature $T$ and have measured (d.c.)
conductivities
at frequencies, $\omega$, which easily satisfy $\hbar \omega \ll k_B T$.
In contrast, all of the theoretical analytical~\cite{fgg,cha,mpaf,wz,kz,fz,ih2}
and exact diagonalization~\cite{runge} work
has computed the universal conductivity at $T=0$, which is equivalent to the
regime $\hbar \omega \gg k_B T$. 
Numerical Monte Carlo work has also measured the universal 
conductivity~\cite{cha,swgy,mtu,bls,wsgy}, but it involves an analytical
continuation from imaginary Matsubara frequencies $i 2 \pi p k_B T/\hbar$ ($p\geq 1$, integer)
to real frequencies: such a procedure is insensitive to any structure
in the conductivity at $\hbar \omega \ll k_B T$, and so these measurements 
are effectively also in the regime $\hbar \omega \gg k_B T$. This discrepancy,
between theory and experiment, in the orders of limit of $\omega \rightarrow 0$
and $T \rightarrow 0$ was noted by Cha~{\em et al.}~\cite{cha} and 
Wallin~{\em et al.}~\cite{wsgy}, but they asserted that the
conductivity approaches a
universal value, at low temperatures and frequencies,
which is {\em independent} of
the value of the ratio $\hbar \omega / k_B T$.

In this paper, we shall argue that the critical conductivity is in fact {\em not}
independent of $\hbar \omega /k_B T$, and is instead a highly non-trivial, but universal,
function of $\hbar \omega / k_B T$; this shall be explicitly demonstrated
in a computation of the crossover function in a model of disorder-free bosons.
We shall show that the physics of the $\hbar
\omega \gg k_B T$
and $\hbar \omega \ll k_B T$ regimes are quite distinct, and shall
provide, for the first
time, a model computation of the universal properties in the
limit $\hbar \omega \ll k_B T$ relevant to d.c. transport measurements. 
We hope that our results will stimulate experimental measurements
of the a.c. conductivity at frequencies which can explore the 
crossover at $\hbar \omega \sim k_B T$.

Our arguments rely on the physical picture of the $T>0$ dynamics at quantum-critical
points developed in the context of quantum antiferromagnets~\cite{sy,csy,statphys,s}.
It was argued in these works that the order parameter dynamics is {\em relaxational\/},
with a relaxation rate $\sim k_B T/\hbar$.  
Using the orientation of the order parameter
to define a phase, we then obtain a phase relaxation
rate $1/\tau_{\varphi} \sim k_B T/\hbar$, 
and this rate is analogous to the 
phase relaxation rate of disordered metals~\cite{aa} 
(the latter rate is much smaller
than $k_B T /\hbar$ in the disordered metal~\cite{aa}, 
but becomes of order $k_B T / \hbar$
at the metal insulator transition--see Section~\ref{ac}).
At time scales of order $\tau_{\varphi}$, the relaxational dynamics cannot
be described by effective classical models of the types discussed in 
Ref~\onlinecite{halphoh}, as a necessary condition
for classical behavior is that the relaxation rate be much smaller than
$ k_B T /\hbar$: the
dynamics was therefore dubbed ``quantum relaxational''.
Related ideas were applied to charge transport in 
the insightful article by Sondhi~{\em et al.}~\cite{shahar}.
Dynamic order parameter fluctuations also carry charge, and therefore 
inelastic collisions between thermally excited charge-carrying
excitations will lead 
to a transport relaxation time $\tau_{\rm tr}$. As the typical
energy exchanged in a collision is $k_B T$, $\tau_{\rm tr}$ is
 also of order $\tau_{\varphi}$,
and therefore~\cite{varma}
\begin{equation}
\frac{1}{\tau_{\rm tr}} \sim \frac{k_B T}{\hbar}.
\label{tautr}
\end{equation}
The missing coefficient in (\ref{tautr}) is a universal number whose value will
depend upon the precise definition of $1/\tau_{\rm tr}$. It is perhaps worth noting
explicitly here that (\ref{tautr}) holds for all values of the dynamic 
exponent $z$. (A scenario under
which (\ref{tautr}) could be violated is discussed in Section~\ref{dii}).
The collisions leading to (\ref{tautr})
conserve total energy and, in the case of translationally invariant systems, total 
momentum. Note that conservation of total momentum should not be confused
with conservation of the total {\em charge current}---the latter is not conserved in
any of the systems of interest here, even in the continuum scaling limit (the
total {\em charge density} is, of course, always conserved).

Now general considerations~\cite{kb} suggest that there are two qualitatively different
regimes of charge transport at non-zero frequencies
\begin{itemize}
\item $\omega \tau_{\rm tr} \ll 1$: the hydrodynamic, incoherent,
collision-dominated regime,
where charge transport is controlled by repeated, inelastic scatterings between
pre-existing thermally, excited carriers; the conductivity should exhibit
a `Drude' peak as a function of frequency.
\item $\omega \tau_{\rm tr} \gg 1$: the high frequency, phase-coherent, collisionless regime, 
where the excitations created by the external perturbation are solely 
responsible for transport, and collisions with thermally excited carriers can be neglected.
\end{itemize}
Essentially all previous theoretical analyses have been restricted to the collisionless
regime.

We can sharpen our description of the crossovers in the 
vicinity of $\omega \sim 1/\tau_{\rm tr}$ by expressing them in universal scaling forms.
We will consider systems where the order parameter quanta carry `charge' $Q$.
For the superfluid-insulator
transition of bosons, we have $Q=2e$ and we measure charge transport by the dynamic
conductivity $\sigma ( \omega)$.
For the N\'{e}el-paramagnet transition in 
antiferromagnets, we have $Q = g \mu_B$ ($\mu_B$ is the Bohr magneton and $g$ is
gyromagnetic ratio) we and measure spin transport by the `spin conductivity'
(also denoted by $\sigma ( \omega )$) which determines the spin current in
response to a uniform gradient in an external magnetic field.
In $d$ spatial dimensions
the dynamic 
conductivity $\sigma$ obeys at the quantum-critical 
coupling
\begin{equation}
\sigma ( \omega) = \frac{Q^2}{\hbar} \left( \frac{k_B T}{\hbar c} \right)^{(d-2)/z}
\Sigma \left( \frac{\hbar \omega}{k_B T} \right)
\label{scaling}
\end{equation}
where $z$ is the dynamic critical exponent, 
$c$ is a non-universal microscopically determined quantity with the dimensions
of $(\mbox{length})^z (\mbox{time})^{-1}$ (for $z=1$, $c$ is a velocity), 
and $\Sigma ( \overline{\omega} ) $ is a universal scaling function (note 
that we are using $\hbar$ rather than $h$ to define the scale
of the conductivity).
We note again that the dependence
on a universal function of $\hbar \omega / k_B T$, with no arbitrary
frequency scale factors, holds for all values of the dynamic exponent
$z$~\cite{sy}.
Scaling  as a function of $\hbar \omega / k_B T$ was also
noted recently by Sondhi~{\em et al.}~\cite{shahar}, but they used it only
to establish the frequency dependence of the conductivity away from 
the critical coupling, with an eye to understanding recent dynamical
conductivity measurements in the quantum Hall system~\cite{engel} (we
will comment on these measurements in Section~\ref{ac}). 
Indeed, their scaling forms are consistent
with a conductivity which is independent of $\hbar \omega/k_B T$
right at the critical point, and hence with implicit or explicit assumptions in
earlier theoretical 
results~\cite{fgg,cha,mpaf,wz,kz,fz,ih2,runge,swgy,mtu,bls,wsgy}.
 One of our new points here is that there is 
a non-trivial dependence on $\hbar \omega / k_B T$ already {\em at} the
critical coupling, and that this dependence means that previous 
analyses of the universal conductivity at the critical coupling either 
did not compute~\cite{fgg,cha,mpaf,wz,kz,fz,ih2,runge}, or were not 
particularly sensitive to~\cite{swgy,mtu,bls,wsgy}, the value of the 
d.c. conductivity.

\begin{figure}
\epsfxsize=5.5in
\centerline{\epsffile{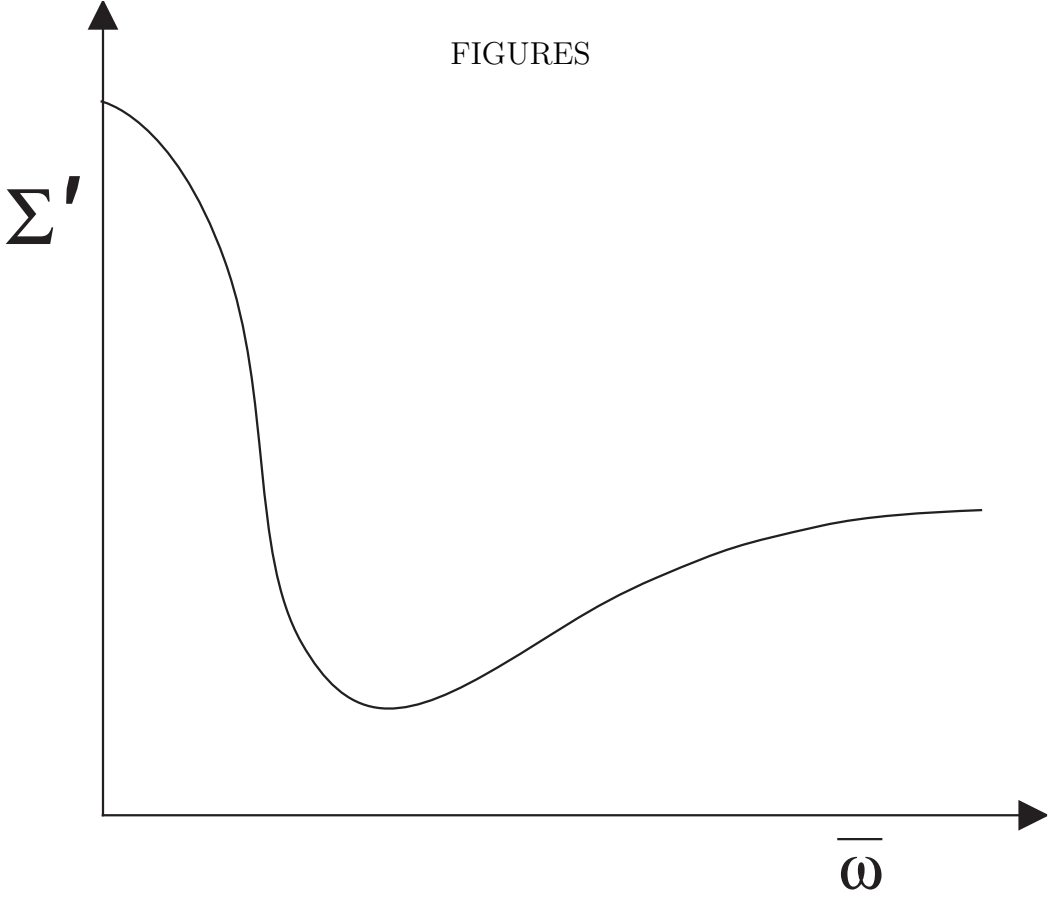}}
\vspace{0.2in}
\caption{A sketch of the expected form of the real part, $\Sigma'$,
of the universal scaling function
$\Sigma$ appearing in the scaling form (\protect\ref{scaling})
for the conductivity, as a function of $\overline{\omega} = \hbar \omega / k_B T$.
There is a Drude-like peak from the inelastic scatterings between thermally excited
carriers at $\overline{\omega}$ of order unity. At larger $\overline{\omega}$, there is 
a crossover to the collisionless regime where
$\Sigma' \sim \overline{\omega}^{(d-2)/z}$ as $\overline{\omega} \rightarrow \infty$.
}
\label{fig1}
\end{figure}

\begin{figure}
\epsfxsize=5.5in
\centerline{\epsffile{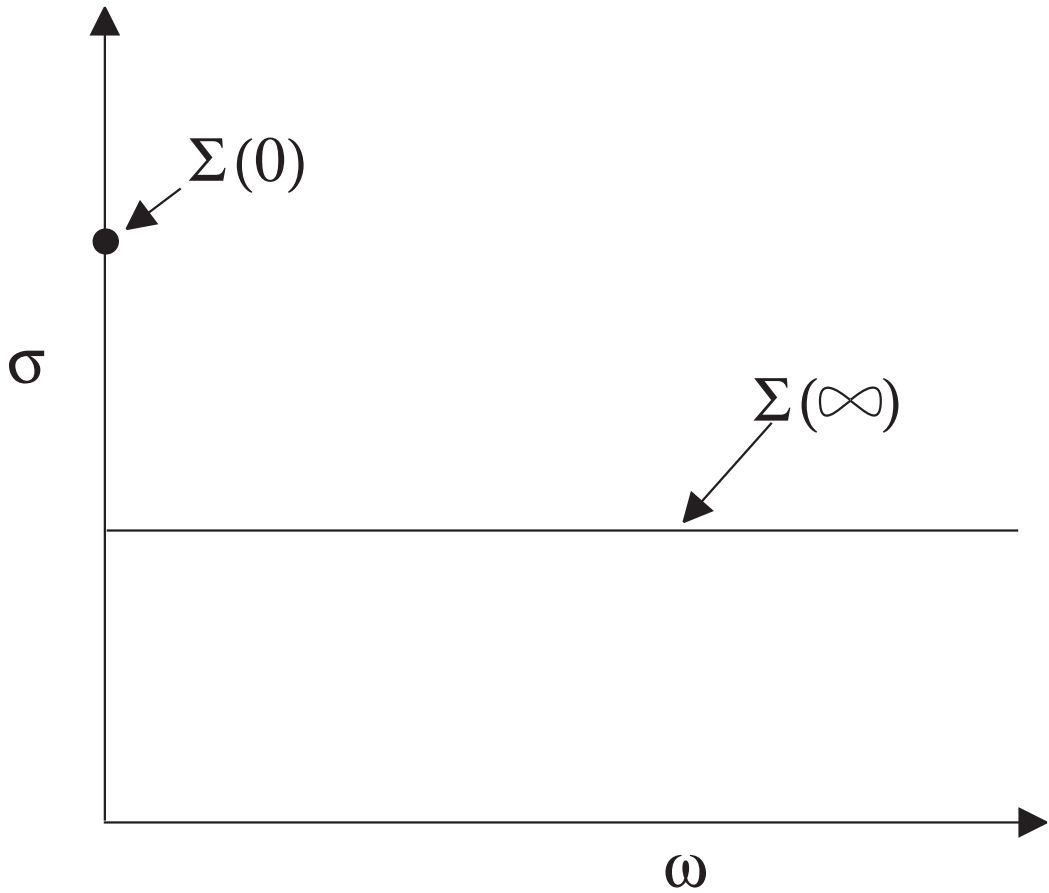}}
\vspace{0.2in}
\caption{Universal form of the conductivity
$\sigma ( \omega , T \rightarrow 0 )$
in $d=2$; the vertical scale is measured in units of $\hbar/Q^2$. 
Only the $\omega =0$ value is given by the universal number 
$\Sigma ( 0 )$. For all $\omega > 0$, $(\hbar/Q^2) \sigma = \Sigma ( \infty )$.
$Q$ is the `charge' of the order parameter: for the superfluid-insulator transition
$Q=2e$, while for quantum antiferromagnets $Q = g \mu_B$.}
\label{fig2}
\end{figure}

Let us now turn to the expected dependence of the conductivity on
the $\hbar \omega / k_B T$ as expressed in the universal function 
$\Sigma ( \overline{\omega} )$. The discussion here will apply
to realistic systems at their critical point for $\hbar \omega$
and $k_B T$ much less than a non-critical microscopic energy scale
{\em e.g.} the repulsion energy $U$ between two bosons on the same
site of a lattice model.
The physical arguments below (\ref{tautr})
have been recast in a qualitative sketch of the real part of the function
$\Sigma ( \overline{\omega})$ in Fig~\ref{fig1}. 
The d.c. conductivity is determined
by the value of the real universal number $\Sigma (0)$. 
At small $\overline{\omega}$ there is a Drude-like peak coming from
the energy-exchanging collisions among thermally excited carriers. At larger
frequencies there is crossover to transport by particle-hole pairs
created by the external source. 
As $\overline{\omega} \rightarrow \infty$ we expect 
that $\Sigma ( \overline{\omega} ) \sim (- i \overline{\omega})^{(d-2)/z}$ so that $\sigma$
becomes independent of $T$ in the collisionless regime. In $d=2$, $\Sigma ( \infty)$
is a real, finite, universal number determining the high frequency conductivity; 
$\Sigma ( \infty )$ was the number computed in earlier 
analyses~\cite{fgg,cha,mpaf,wz,kz,fz,ih2,runge},
and not the d.c. conductivity which is given by $\Sigma ( 0)$.
This is illustrated in Fig~\ref{fig2} which plots the form of
$(\hbar/Q^2) \sigma ( \omega, T \rightarrow 0)$ in $d=2$: its value at $\omega =0$ is
given by $\Sigma (0)$, while for all $\omega > 0$ it equals $\Sigma ( \infty )$.
It is likely, although not established, that $\Sigma (0) > \Sigma (\infty)$.
Note the difference from Fermi liquid theory, where the Drude peak becomes
a delta function with non-zero weight as $T \rightarrow 0$. In the present
situation,
the weight in the Drude-like peak vanishes like $\sim T^{(d+z-2)/z}$ as $T \rightarrow 0$,
and in $d=2$ and the Drude-like peak
reduces to the {\em single} point $\omega = 0$ where the conductivity is
given by $\Sigma ( 0)$. The scaling properties of the $d=1$, $z=1$ case are 
similar to those of a Fermi liquid.

There is a certain critical phenomena/scaling perspective in which the
result (\ref{scaling}), and its implications discussed above, 
may seem quite natural, and even
somewhat `trivial' as they follow directly from the fact that 
$\omega$ and $T$ have the same scaling dimension. Nevertheless, 
its importance has been overlooked in essentially all previous work.
This is probably because there are complementary perspectives, more common
among investigators in this field, from which 
(\ref{scaling}) implies physics that is
surprising and even somewhat radical. 
In particular, current ideas in 
quantum transport theory~\cite{alw}
and dissipative quantum mechanics~\cite{uw} can lead one to rather different
picture, as we now itemize explicitly:
\begin{itemize}
\item
There have been a number of previous situations
in which charge transport properties have been found to be universally related
to the quantum unit of conductance, $e^2 / h$; these include the 
quantized Landauer conductance of ballistic transport in one-dimensional
wires, and the universal conductance fluctuations of mesoscopic metals~\cite{alw,ucf}. 
However in all
previous cases, these universal properties have arisen in a ``phase-coherent'' regime,
{\em i.e.\/} they are associated with physics at scales shorter than the mean
distance between inelastic scattering events between the carriers. For the
case of a $d=2$ quantum critical point discussed above, the 
universal number $\Sigma ( \infty )$
is associated with quantum coherent transport, and is therefore 
the analog of these earlier results. In contrast, the value of $\Sigma (0)$ is controlled by
repeated inelastic scattering events, and therefore the d.c. transport
is clearly in what would traditionally
be identified as the ``incoherent'' regime. Nevertheless, we have argued above that
$\Sigma (0) $ is a universal number, and remarkably, the d.c. conductance remains
universally related to $e^2 / h$.
\item
The community has gained much intuition on
the non-zero temperature 
transport properties of interacting quantum systems from recent studies of  
dissipative quantum mechanics and a number of related
quantum-{\em impurity} problems~\cite{uw,fls,lss,kf}. The scaling properties of such models
are given by the theory of {\em boundary\/} critical phenomema, in contrast to the {\em
bulk\/} critical  phenomena of interest in this paper. 
We discuss the transport properties of a class of
boundary problems in Appendix~\ref{dqm}:
the leading term in the transport coefficient is found to be independent of $\omega / T$,
and dependence on $\omega / T$ arises only upon consideration of subleading
terms at low $T$. However, as we argue in Appendix~\ref{dqm}, this behavior
is understood by the fact that the fixed point controlling
the low $T$ behavior is simply a {\em free field\/} theory, and 
$\omega /T$ dependence arises only upon considering the leading irrelevant operator.
In contrast, the bulk theories of interest here differ in a crucial respect:
they have {\em interacting} critical theories
and therefore, we argue, contain $\omega/T$ dependence already in the leading
term, before the inclusion of any irrelevant operators.
\item 
Scaling as a function of $\omega/T$ does not hold for the Anderson localization
transition of non-interacting electrons. The frequency obeys conventional scaling
with dimension $z$, but the behavior of temperature is rather different and
non-universal. Analytic theories for such transition are available only 
for $d>2$, and in these the primary effect of temperature is in the non-universal,
superlinear in $T$ dependence of the phase-breaking rate which acts like
a finite-size-like infrared cutoff to the critical properties. See also Section~\ref{dii}.
In contrast, the models of interest here have interactions, and inelastic scattering is central
to understanding their universal critical properties; in some respects scaling
with respect to temperature is simpler, as its naive scaling dimension of $z$ is now valid.
\end{itemize}

In this paper we will provide explicit results for the crossover function
$\Sigma ( \overline{\omega})$ in a simple, disorder-free field theoretic model
for the superfluid insulator transition that was introduced by 
Cha~{\em et al.}~\cite{cha}.
Near the quantum-critical point, this model becomes equivalent to the familiar
particle-hole symmetric 
$\phi^4$ field theory with the effective imaginary time ($\tau$) action
\begin{equation}
{\cal S} = \int_0^{\hbar/k_B T} d\tau \int d^d x \left\{
\frac{1}{2} \left[ ( \partial_{\tau} \phi_{\alpha} )^2 + c^2 ( \nabla_x \phi_{\alpha}
)^2 + ( m_{0c}^2 + t_0 ) \phi_{\alpha}^2 \right] 
+ \frac{u_0}{4!} ( \phi_{\alpha}^2 )^2 \right\}.
\label{action}
\end{equation}
Here $\phi_{\alpha}$ is a $n$-component field and the action has $O(n)$ symmetry
(the $O(n)$ index $\alpha$ is implicitly summer over). 
The spatial and temporal gradient terms are both second order, so the action
has a ``Lorentz'' invariance with $c$ the velocity of light, and as a result
the dynamic critical exponent $z=1$.
The bare ``mass'' term has been written as $m_{0c}^2 + t_0$ so that the
$T=0$ quantum critical point is at $t_0 = 0$, and $u_0$ measures the strength 
of the quartic non-linearity.
The superfluid-insulator transition
is described by the case $n=2$ where $\Psi = \phi_1 + i \phi_2$ is the usual
complex superfluid order parameter. We shall also be interested in the case
$n=3$ which applies to quantum-critical points in quantum antiferromagnets~\cite{s}.

It should be noted that the continuum model ${\cal S}$ for $n=2$, with
its double time derivative term,
differs from the usual
continuum action for non-relativistic bosons~\cite{popov} which has
only a single time derivative of a complex scalar field. The total
momentum and charge current of the latter model are proportional to each other.
When these
non-relativistic bosons are placed under the influence of an
effective lattice potential, with an average of an integer
number of bosons per lattice site, then after integrating out certain local high energy
modes, one obtains the continuum model ${\cal S}$ as a low energy effective
theory~\cite{fwgf}. The charge current and momentum operators of ${\cal S}$ are
now no longer simply related to each other as 
${\cal S}$ has excitations with both positive and 
negative charges.
As we will see in Section~\ref{qtrans}, the total momentum is 
proportional to the {\em sum\/} of the currents of the
positive and negative charges, while the total charge current is proportional to 
their {\em difference\/}.
This lack of a direct relationship between the charge current
and the momentum should not seem surprising as lattice effects
were required to obtain ${\cal S}$, and are therefore implicitly
accounted for in the continuum theory.

It was asserted by Cha~{\em et al.}~\cite{cha} that the $T>0$ transport 
properties of ${\cal S}$ 
were ``pathological'' in that there was zero resistance to
charge transport in the continuum 
field theory in (\ref{action}); they suggested that a non-zero resistance
appeared only upon considering additional lattice corrections (beyond those
required to derive the continuum theory ${\cal S}$), in which momentum could
jump in units of reciprocal lattice vectors at scattering events (the
so-called ``umklapp'' scattering events). We shall show here that
this is incorrect. The model ${\cal S}$ has a finite, and 
universal, d.c. resistance at any
$T>0$ already in the scaling, continuum limit.
Indeed, this is already clear from a recent study~\cite{o3} of ${\cal S}$
for the case $d=1$ $n=3$, where a simple and physically transparent argument
obtained the exact (and finite) low temperature value of the spin diffusivity.
Rather than being pathological, 
we claim
that the transport properties of the disorder-free boson model ${\cal S}$ are 
generic, and essentially 
identical in their scaling structure to those of disordered boson systems.
The error in Cha~{\em et al.}~\cite{cha} appears to be due to their ignoring
the presence of independent positive and negative charge excitations,
and the resulting difference between 
the total momentum and the total charge current of ${\cal S}$~\cite{luttinger}.

As $T$ appears in ${\cal S}$ only in the upper limit of the imaginary time integral,
it is clear that the scaling form (\ref{scaling}) is nothing but a standard
finite-size scaling result for observables as a function of ``wavevector'', $\omega$,
and ``inverse size'', $T$. It might then seen that our job is relatively straightforward,
and we merely have to obtain standard finite-size scaling results on the familiar
model ${\cal S}$. This is far from being the case. The point is that these
standard results exist only at imaginary frequencies $i 2 p \pi k_B T /\hbar$ ($p \geq 1$, 
integer),
and we are especially interested in real frequencies $\ll k_B T/\hbar$. More importantly, 
it has
been shown~\cite{sy,csy,s}, that the operations of analytic continuation 
and expansion in $1/n$ or $\epsilon=3-d$ (which are the only non-numeric tools
for analyzing the critical point of ${\cal S}$) {\em do not commute}.

The proper tool for analyzing
transport at the critical point of ${\cal S}$ 
is a quantum Boltzmann equation (QBE) for the charge carriers. 
In general, solution of such a QBE is a dauntingly difficult
task, but we shall find that it is possible to reduce the solution to a simple
linear integral equation in an expansion~\cite{fz,s}
in $\epsilon = 3-d$ (very similar results can be obtained in 
a related analysis of ${\cal S}$
in an expansion in $1/n$~\cite{cha,sy,csy}; this will be described elsewhere, 
and we will only discuss the $n=\infty$ results here.). 
For small $\epsilon$, the non-linearities are weak,
and it becomes possible to give a quasiparticle-like
interpretation to the excitations of ${\cal S}$ at the quantum critical point.
Such an interpretation will be useful in our intuitive understanding, and will
help us use standard methods to simplify the QBE in 
the $\hbar \omega \ll k_B T$ limit.
Although the quasiparticle interpretation fails
for the physical case $\epsilon=1$, we do not expect any qualitative
change in the structure of our results for larger $\epsilon$. The QBE
formulation is quite general, and the quasiparticle representation
is mainly a useful technical tool towards obtaining its numerical solution. 
In a more general context, our approach may considered as an expansion
in powers of the anomalous dimension, $\eta$, which is responsible for
replacing the quasi-particle pole by a continuum at the $T=0$ critical point.
The continuum associated with a non-zero $\eta$ is important only for $\hbar \omega \gg 
k_B T$, while for $\hbar \omega \sim k_B T$ very different thermal damping processes
quench the critical fluctuations~\cite{statphys}, and
are best treated by the QBE. This thermal damping
also acts as an effective infrared cutoff which ensures that no qualitatively new physics
emerges at higher orders in the expansion in powers of $\eta$.

\begin{figure}
\epsfxsize=5.5in
\centerline{\epsffile{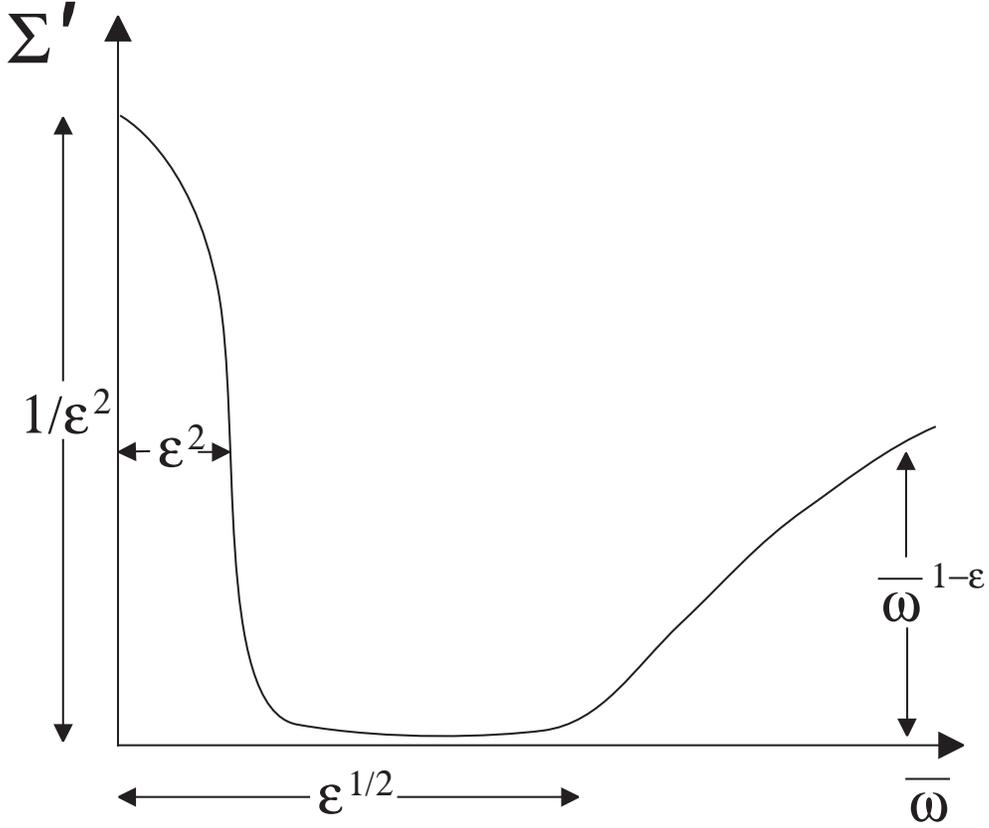}}
\vspace{0.2in}
\caption{Structure of the real part, $\Sigma'$, of the
universal scaling function $\Sigma$ in (\protect\ref{scaling})
for the conductivity at the quantum critical coupling of the model ${\cal S}$ defined
in (\protect\ref{action}). The spatial dimensionality $d=3-\epsilon$, and $\epsilon$
is assumed to be small. As before $\overline{\omega} = \hbar\omega/ k_B T$. 
The Drude peak at small $\overline{\omega}$ has a width of order $\epsilon^2$ and a
height of order $1/\epsilon^2$: this feature of the conductivity is denoted later in the paper
by $\sigma_I$. The collisionless contribution (denoted $\sigma_{II}$ later)
begins at $\overline{\omega}$ of
order $\epsilon^{1/2}$; as $\overline{\omega} \rightarrow \infty$, this contribution
is a number of order unity times $\overline{\omega}^{1-\epsilon}$
}
\label{fig3}
\end{figure}

The basic structure of the dynamical conductivity at the critical point of ${\cal S}$
for small $\epsilon$ is illustrated in Fig~\ref{fig3}. 
Notice the presence of ``boundary layers'' which make the analysis of the
$\epsilon \rightarrow 0$ limit
quite subtle, and is responsible for the non-commutativity of analytic continuation
and the naive $\epsilon$ expansion noted earlier. 
The hydrodynamic regime of the conductivity
(denoted later in the paper by $\sigma_I$) is represented by a Drude peak of 
width in frequency $\omega \sim \epsilon^2 k_B T/\hbar$ and has a height of order $1/\epsilon^2$.
In particular, the d.c. conductivity is determined by the universal number $\Sigma (0)$, for
which we find for $n=2$
\begin{equation}
\Sigma (0) = \frac{0.1650}{\epsilon^2},
\label{i1}
\end{equation}
to leading order in $\epsilon$; the structure of the higher-order corrections
to (\ref{i1}) is quite complex, and was generally discussed in Ref~\onlinecite{s}.
Determination of the coefficient in (\ref{i1})
required the numerical solution of a QBE, and the uncertainty
in the numerics is believed to be restricted to the fourth decimal place.
In $d=2$, (\ref{i1}) implies a universal d.c. conductivity 
\begin{equation}
\sigma (0) = 
2 \pi \Sigma (0) \frac{4e^2}{h} \approx 1.037 \frac{4e^2}{h}
\label{i1b}
\end{equation}
at the superfluid-insulator transition. 
This result is remarkably close to the 
self-dual value $4 e^2 /h$~\cite{mpaf,wz}, and to the results of many experiments~\cite{lg};
we will comment further on this in Section~\ref{duality}.

There is a clean separation between the hydrodynamic and collisionless regimes
of $\sigma$: the latter does not begin until $\omega \sim \epsilon^{1/2} 
k_B T / \hbar$ (Fig~\ref{fig3}). 
The $T=0$ collisionless transport is characterized by $\Sigma ( 
\overline{\omega} \rightarrow \infty)$ for which~\cite{fz} we have at $n=2,3$
\begin{equation}
\Sigma' ( \overline{\omega} \rightarrow \infty ) = 
\frac{2^{1-2d} \pi^{1-d/2}}{d \Gamma (d/2)} \left( 
1 + {\cal O} (\epsilon^2 ) 
\right) \overline{\omega}^{1-\epsilon} .
\label{i1aa}
\end{equation}
As noted earlier, for $\epsilon=1$ ($d=2$) $\Sigma ( \infty)$ is a pure number,
but notice that it bears no relationship to $\Sigma (0)$; indeed $\Sigma ( \infty)$ is
of order unity as $\epsilon \rightarrow 0$, and so
these two quantities
are of distinct orders in $\epsilon$.
For the superfluid-insulator transition in $d=2$, (\ref{i1aa}) gives a high frequency
conductivity of $0.3927 \times (4 e^2 /h)$, which is in rough agreement with
other analyses in the collisionless regime~\cite{fz}.

The body of the remainder of the paper is devoted to obtaining the above properties of
the model ${\cal S}$. Readers not interested in calculations specific to 
the model ${\cal S}$ should now go
directly to the concluding Section~\ref{conc} where we discuss implications
of our results for a number of experimental systems. 
In Section~\ref{sec:kubo} we will obtain one loop results for the transport
properties of ${\cal S}$ using the familiar Kubo formalism. Then
Section~\ref{qtrans} will include two-loop effects using a 
quantum transport analysis needed to describe the hydrodynamic
regime, and obtain (\ref{i1}).

In all of Sections~\ref{sec:kubo},~\ref{qtrans}, and the appendices
we will work in units in which
$\hbar = k_B = c = 1$; we will reinstate these constants in Section~\ref{conc}.

\section{One loop results from the Kubo formula}
\label{sec:kubo}
We will begin our analysis of the transport properties of ${\cal S}$ by examining the
results of a direct evaluation from the Kubo formula~\cite{cha,fz}, but now working
at $T>0$. The physical interpretation of the results will motivate an analysis
using a quantum Boltzmann equation~\cite{kb,keldysh,pd} which will be carried 
out in subsequent sections.

The standard Kubo formula relates the conductivity to a two-point correlator
of the conserved $O(n)$ current. We introduce an external vector potential
${\bf A}$ associated with the $O(n)$ generator which rotates $\phi_{\alpha}$
in the $1,2$ plane; the spatial gradient term in ${\cal S}$ then undergoes
the mapping
\begin{equation}
\sum_{\alpha=1}^{n} \left( \nabla_x \phi_{\alpha} \right)^2 \rightarrow
\left( \nabla_x \phi_{1} - Q {\bf A} \phi_2 \right)^2 +
\left( \nabla_x \phi_{2} + Q {\bf A} \phi_1 \right)^2 +
\sum_{\alpha=3}^{n} \left( \nabla_x \phi_{\alpha} \right)^2.
\label{vecpot}
\end{equation}
The associated $O(n)$ current is then $\delta {\cal S}/\delta {\bf A}$.
We evaluate its two-point correlator using standard
diagrammatic perturbation theory to first order in $u_0$ in the expansion in 
$\epsilon$, or to leading order in the large $n$ expansion (in which $u_0 \sim 1/n$). 
The first order
vertex correction vanishes because the interaction is momentum independent
and the result in both cases is given simply by~\cite{cha,fz} (recall we are using units
here in which $\hbar = k_B = c = 1$)
\begin{equation}
\sigma ( i \omega_n ) = - \frac{2 Q^2}{\omega_n}
T \sum_{\epsilon_n} \int \frac{d^d k}{(2 \pi)^d}
\left[
\frac{2 k_x^2}{(\epsilon_n^2 + k^2 + m^2)((\epsilon_n+\omega_n)^2 + k^2 + m^2)}
 - \frac{1}{\epsilon_n^2 + k^2 + m^2} \right].
\label{k1}
\end{equation}
The first term is the `paramagnetic' contribution, while the second
is the `diamagnetic' term proportional to the density~\cite{mahan}.
Here $\epsilon_n, \omega_n$ are Matsubara frequencies, $k_x$ is the $x$ component
of the $d$ dimensional momentum ${\bf k}$,  and $k = |{\bf k}|$. The ``mass'' $m$
in the propagators is computed in Appendix~\ref{susc} (where it is referred to as
$m(T)$) using the $\epsilon = 3-d$ expansion developed in Ref~\onlinecite{s}, and 
depends universally
upon $T$ and an energy scale measuring the deviation of the ground state
from the critical ground state. At the 
critical coupling, to leading order in $\epsilon$,
\begin{equation}
m^2 = \epsilon \left( \frac{n+2}{n+8} \right) \frac{2 \pi^2 T^2}{3}~~~~~~~~~~
\mbox{at $t_0 = 0$}
\label{k1a}
\end{equation}
The result (\ref{k1}) holds for all $n$, and there is no $n$ dependence
at this order in the $\epsilon$ expansion, other than that through $m$.
The large $n$ expansion has an identical structure at $n=\infty$, the
only difference being in the value of $m$. Detailed universal expressions for
$m$ were given in Ref.~\cite{csy} in $d=2$; at the critical coupling, the analog
of (\ref{k1a}) at $n=\infty$ is
\begin{equation}
m = 2 \ln \left( \frac{\sqrt{5} + 1}{2} \right) T
~~~~~~~~~~
\mbox{at $t_0 = 0$, $d=2$}
\label{k1b}
\end{equation}
The remaining analysis of this section will apply both to the $\epsilon$ and $1/n$
expansions, the only difference being in the values of $m$ given above.

Now insert $1 = \partial k_x/\partial k_x$ in front of the diamagnetic term
in (\ref{k1}) and integrate by parts. The surface terms vanish 
in dimensional or lattice regularization, and the expression for the conductivity becomes
\begin{equation}
\sigma ( i \omega_n ) = -\frac{2Q^2}{\omega_n}
T \sum_{\epsilon_n} \int \frac{d^d k}{(2 \pi)^d}
\frac{2 k_x^2}{\epsilon_n^2 + k^2 + m^2} \left[\frac{1}{(\epsilon_n+\omega_n)^2 + k^2 + m^2}
 - \frac{1}{\epsilon_n^2 + k^2 + m^2} \right].
\label{k2}
\end{equation}
We now evaluate the summation over Matsubara frequencies, analytically continue to real
frequencies. The resulting $\sigma ( \omega)$ is complex, and we decompose
it into its real and imaginary parts $\sigma (\omega) = \sigma'(\omega) + i \sigma''(\omega)$.
We will only present results for the real part $\sigma'(\omega)$,
and the imaginary part
$\sigma''(\omega)$ can be obtained via the standard dispersion relation.

We find that the result for $\sigma'(\omega)$ has two distinct contributions
of very different physical origin. We separate these by writing
\begin{equation}
\sigma'(\omega) = \sigma'_{I} (\omega) + \sigma'_{II} ( \omega ).
\label{k3}
\end{equation}
The first part, $\sigma'_{I} ( \omega)$, is a delta function
at zero frequency: 
\begin{equation}
\sigma'_{I} ( \omega) = 2\pi Q^2 \delta ( \omega ) \int \frac{d^d k}{(2 \pi)^d}
\frac{k_x^2}{\varepsilon_k^2} \left( - \frac{\partial n ( \varepsilon_k )}{\partial
\varepsilon_k} \right),
\label{k4}
\end{equation}
where $n(\varepsilon)$ is the Bose function
\begin{equation}
n ( \varepsilon ) = \frac{1}{e^{\varepsilon/T} - 1},
\end{equation}
and the excitations have the energy momentum relation
\begin{equation}
\varepsilon_k \equiv \sqrt{ k^2 + m^2}
\label{k5}
\end{equation}
We will discuss the physical meaning of the delta function in (\ref{k4}) below, and obtain
a separate and more physical derivation of its weight in Section~\ref{clesstrans}. 
The second part, $\sigma'_{II} ( \omega )$ is a continuum above a threshold
frequency of $2 m$:
\begin{eqnarray}
\sigma'_{II} ( \omega ) &=& \pi Q^2 \int \frac{d^d k}{(2 \pi)^d}
\frac{ k_x^2}{2 \varepsilon_k^3} ( 1 + 2 n( \varepsilon_k )) \delta ( |\omega| - 2
\varepsilon_k ) \nonumber \\
&=& \frac{\pi Q^2 S_d}{d} \theta( |\omega| - 2 m) \left( \frac{\omega^2 - 4 m^2}{4 \omega^2}
\right)^{d/2} \left[ 1 + 2 n(\omega/2) \right] |\omega|^{d-2},
\label{k5a}
\end{eqnarray}
where $S_d = 2/(\Gamma (d/2) (4 \pi)^{d/2})$
is a standard phase space factor.

\begin{figure}
\epsfxsize=5.5in
\centerline{\epsffile{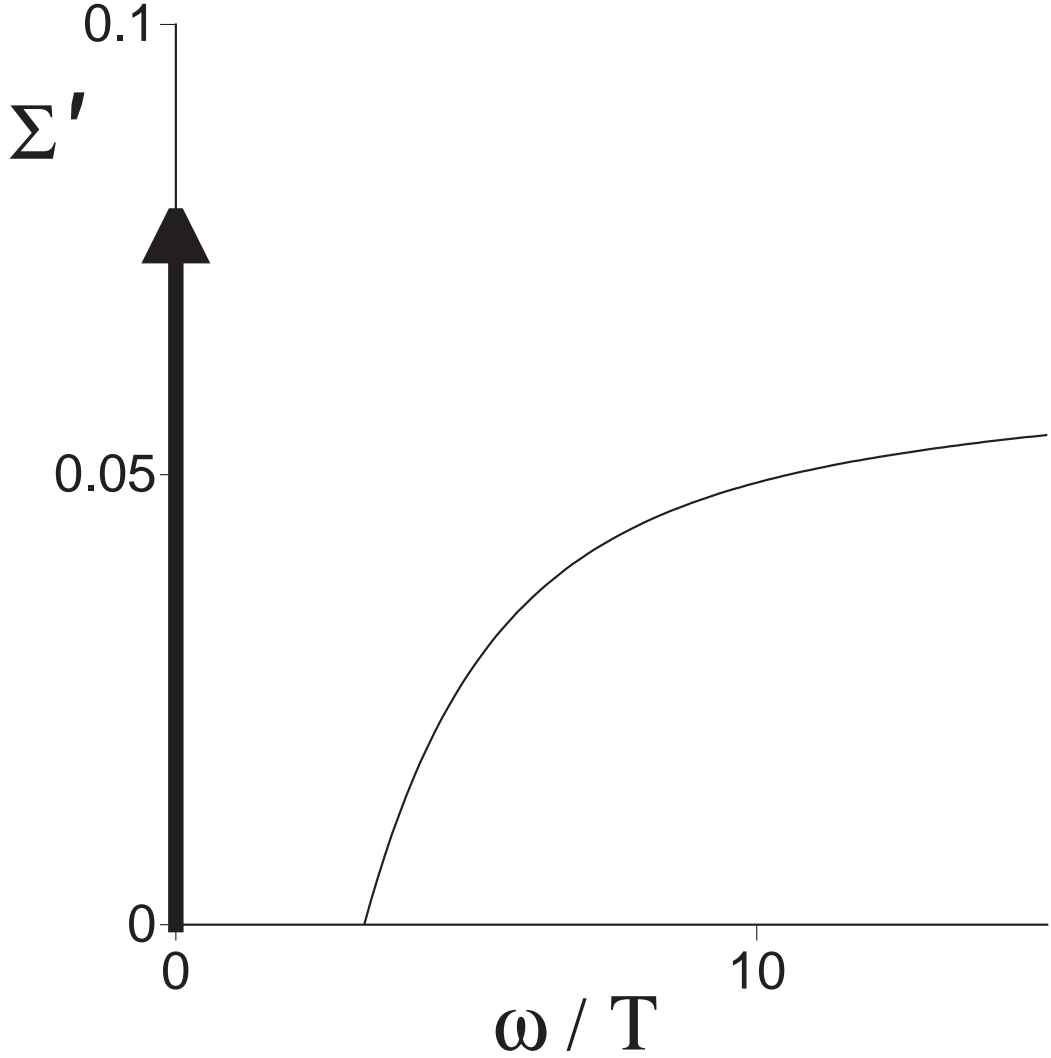}}
\vspace{0.2in}
\caption{The real part, $\Sigma'$, of the 
universal scaling function $\Sigma$ (see (\protect\ref{scaling}))
for the conductivity at the quantum critical coupling of the model ${\cal S}$,
 correct to first order in $\epsilon=3-d$. The numerical
values are obtained from (\protect\ref{k1a}) and (\protect\ref{k5a}) 
with $d=2$ ($\epsilon = 1$). 
There is a delta function precisely
at $\omega/T = 0$ represented by the heavy arrow: the weight of this delta function
is given in (\protect\ref{k4}) and (\protect\ref{k6}). The delta function
contributes to $\sigma_I$, and the higher frequency continuum to $\sigma_{II}$
}
\label{fig4}
\end{figure}

At the critical point, $t_0 = 0$, it can be verified that the above results
for $\sigma (\omega )$ obey the scaling form (\ref{scaling}) with $z=1$; explicit results
for the function $\Sigma ( \overline{\omega})$ will appear below and are
sketched in Fig~\ref{fig4}.

We now discuss the physical and scaling properties of the two components of the conductivity
in turn:

\subsubsection{$\sigma_I$}

This is a zero frequency delta function, and is present only for $T>0$.
Clearly, it must be interpreted as the contribution of thermal
excitations which propagate ballistically. Indeed, to first 
order in $\epsilon$~\cite{s}, or at $n=\infty$~\cite{csy},
the excitations are simply undamped particles and holes with an infinite
lifetime and
energy momentum
relation $\varepsilon_k$. It is necessary to go to second order in $\epsilon$, 
or to first order
in $1/n$, to include
collisions which will give the quasiparticles a finite lifetime. We will show in
Section~\ref{qtrans} that these collisions also broaden the delta function in $\sigma_I$.
The magnitude of the broadening is expected to be determined by the inverse
lifetime of the quasiparticles; at the critical
point, this inverse lifetime is of order $\epsilon^2 T$~\cite{s} in the $\epsilon$
expansion, or of order $T/n$~\cite{csy} in the large $n$ theory.
The typical energy of a quasiparticle at the critical point is of order $T$, and
so the quasiparticles are well-defined, at least within the $\epsilon$ or $1/n$ 
expansion.
Notice, however, that the quasiparticle interpretation breaks down at the physically
important value of $\epsilon = 1$, $n=2$. The discussion in Section~\ref{qtrans} will take
place within the context of the $\epsilon$ expansion, and will use the 
quasiparticle interpretation and the arsenal of powerful techniques 
available~\cite{kb,keldysh,pd} to describe their quantum
transport.

The expression (\ref{k4}) is valid everywhere in the normal phase, but here we
evaluate it explicitly only at $t_0 = 0$. Consider first the $\epsilon$ expansion.
The coefficient of the delta function is
a function of the ratio $m/T$, but notice from (\ref{k1a}) that $m \ll T$ at 
for small $\epsilon$. Evaluating (\ref{k4}) in this limit we find for $\epsilon$ small
\begin{eqnarray}
\sigma'_I ( \omega ) &=& 2 \pi Q^2 T^{d-1} \delta ( \omega ) 
\left[ \frac{1}{18}	 
- \frac{m}{8 \pi T} + \ldots
\right] \nonumber \\
&=& 2 \pi Q^2 T^{d-1} \delta ( \omega ) 
\left[ \frac{1}{18} 
- \frac{\sqrt{\epsilon}}{8} \left( \frac{2(n+2)}{3(n+8)} \right)^{1/2}
+ \ldots \right]
\label{k6}
\end{eqnarray}
Actually the expression (\ref{k4}) is good to order $\epsilon$ but we have refrained from
displaying the next term as it is rather lengthy. The first term in (\ref{k6}) is obtained
by evaluating (\ref{k4}) at $m=0$, $d=3$; the second term is from an integral dominated
by small $k \sim m \ll T$ and hence the Bose function can be replaced by its classical limit.
An important point to note is that the current carried by the thermally excited carriers
is dominated in the leading term of (\ref{k6}) by momenta $k \sim T \gg m$. This will be
useful to us in the analysis of collisions in Section~\ref{qtrans} where we will simply
be able to set $m=0$ to obtain the leading term.

Turning to the large $n$ theory, we evaluate (\ref{k4}) in $d=2$ using the value of
$m$ in (\ref{k1b}), and obtain after an integration by parts and rescaling of 
variables
\begin{eqnarray}
\sigma'_I ( \omega ) &=& \frac{Q^2 T}{2} \delta ( \omega ) 
\left[ \int_{\Theta}^{\infty} d \varepsilon \left( 1 + \frac{\Theta^2}{\varepsilon^2} \right)
\frac{1}{e^{\varepsilon} - 1} \right] \nonumber \\
&=& \frac{Q^2 T}{2} \delta ( \omega ) \times 0.689402548116632\ldots
\label{k7}
\end{eqnarray}
where $\Theta = 2 \ln ( ( \sqrt{5} + 1)/2)$ is a number which plays a central role in the
large $n$ theory~\cite{csy,poly}.
Notice that as $m \sim T$, we have now been unable to approximate $\varepsilon_k \approx k$
to get the leading result, as was done in the $\epsilon$ expansion. 

An interesting numerical property of the above results in $d=2$ is worth noting explicitly.
The spectral weight of the delta function to leading order in the $\epsilon$ expansion
is, from (\ref{k6}), $Q^2 T$ times $\pi/9 = 0.3491\ldots$ (recall that this number
was obtained by evaluating a momentum space integral in $d=3$). The same quantity
at $n=\infty$, from (\ref{k7}), is $Q^2 T$ times $0.3447\ldots$ (obtained now by
evaluating a different momentum space integral in $d=2$), which is remarkably close.
Later in Section~\ref{qtrans}, 
we will consider broadening of the delta function in the $\epsilon$ expansion,
and we will work in the approximation in which the spectral weight is 
$Q^2 T$ times $\pi/9$ (see Eqn (\ref{inteq4}) later);
the present numerical ``coincidence'' suggests that the numerical values of the
leading order $\epsilon$ result are quite accurate.

\subsubsection{$\sigma_{II}$}

This is the continuum contribution to $\sigma$ which vanishes for $\omega < 2m$. At this
order in $\epsilon$ ($1/n$) there is a sharp threshold at $\omega = 2m$ but we expect that
this singularity will be rounded out when collisions are included at order 
$\epsilon^2$ ($1/n$): we
will not describe this rounding out it in this paper, however. 
Although they have a  strong effect
at the threshold, collisions are not expected to significantly modify the form 
of $\sigma'_{II} ( \omega ) $ at higher frequencies where the transport is
predominantly collisionless. In particular, the $\omega \rightarrow \infty$ limit
is precisely the $T=0$ result obtained earlier~\cite{fz}
\begin{equation}
\sigma'_{II} ( \omega \rightarrow \infty ) = \frac{\pi Q^2 S_d}{ 2^d d} |\omega|^{d-2}
\end{equation}

\section{Quantum transport equations}
\label{qtrans}
The general analysis of higher order corrections to $\sigma$ is quite complex,
and so we will confine ourselves in this section to the answer to a single question: 
how does the $\delta ( \omega)$
term in $\sigma'_I ( \omega )$ broaden ? 

We will address this question exclusively
in the context of
the $\epsilon$ expansion, and generalization to the  $1/n$
expansion will be discussed elsewhere.
In principle, the answer can be obtained
by including the ${\cal O} ( \epsilon^2 )$ correction to the self energy and accounting
for the associated infinite-ladder vertex corrections~\cite{mahan}. 
However this method is quite inconvenient
as it does not allow for easy separation of the distinct phenomena in different frequency
regimes. Instead, we shall use the (in principle) equivalent~\cite{mahan} quantum transport 
formalism of Kadanoff and Baym~\cite{kb,keldysh,pd}. The physical content of this formalism is
transparent at all stages, and the approximations necessary to focus on the low frequency
conductivity are readily apparent. In particular, we can drop the terms leading to 
$\sigma_{II} ( \omega )$ at an early stage.

The transport equation is best studied in the Hamiltonian formalism
by casting it in terms 
of the weakly interacting ``normal modes'' .
So, we begin by writing down the Hamiltonian associated with ${\cal S}$
(reminder---we are using units in which $\hbar = k_B = c = 1$):
\begin{equation}
{\cal H} = {\cal H}_0 + {\cal H}_{\rm int} + {\cal H}_{\rm ext}
\end{equation}
The first term, ${\cal H}_0$ is the free particle part of ${\cal S}$,
\begin{equation}
{\cal H}_0 = \frac{1}{2} \int \left[
\pi_{\alpha}^2 + (\nabla_x \phi_{\alpha})^2 + m^2 \phi_{\alpha}^2 \right],
\end{equation}
where $\pi_{\alpha} (\bx,t)$ ($t$ is real time)
is the canonically conjugate momentum to the quantum 
field $\phi_{\alpha} (\bx, t)$ and satisfies the equal-time commutation relations
\begin{equation}
\left[ \phi_{\alpha} (\bx,t) , \pi_{\beta} (\bx',t) \right] = i \delta_{\alpha\beta}
\delta^d ( \bx - \bx' ).
\label{qt1}
\end{equation}
For future convenience,
we have already included the Hartree-Fock correction from the interactions
into the `mass' $m^2$ (these correspond to self-consistently summing all the
one-loop tadpole diagrams, as discussed in Appendix~\ref{susc}). 
The second term, ${\cal H}_{\rm int}$ is the quartic
interaction
\begin{equation}
{\cal H}_{\rm int} = \frac{u_0}{4!} \int d^d x ( \phi_{\alpha}^2 )^2,
\end{equation}
and it is understood that the Hartree-Fock term arising from ${\cal H}_{\rm
int}$ will be omitted. Finally, ${\cal H}_{\rm ext}$ contains the 
coupling to the external space and time-dependent potentials $U^a (\bx, t)$
($a = 1 \ldots n(n-1)/2$)
\begin{equation}
{\cal H}_{\rm ext} =  Q \int d^d x U^a (\bx,t) L^a_{\alpha\beta} \pi_{\alpha}
(\bx,t) \phi_{\beta} (\bx,t).
\end{equation}
Here the $L^a$
are $n\times n$ real, antisymmetric matrices that are $i$ times 
the generators of the Lie algebra of
$O(n)$. The $U^a$ are coupled to the conserved $O(n)$ charge densities~\cite{iz} of
${\cal H}$ + ${\cal H}_{\rm int}$. 
We shall be interested only in the linear response of the current
to the `electric field' ${\bf E}^a = - \vec{\nabla}_x U^a (\bx,t)$, and it will
be assumed below that ${\bf E}^a$ is independent of $\bx$. 
Notice that we are making a gauge choice different from that in (\ref{vecpot}), and coupling
now to the scalar and not the vector potential; this is for convenience, and should
not change the final gauge-invariant results. 
The `charge' current
${\bf J}^a$ is defined by the expectation value~\cite{iz}
\begin{equation}
{\bf J}^a = Q L^a_{\alpha\beta} \left\langle \phi_{\alpha} \vec{\nabla}_x
\phi_{\beta} \right \rangle,
\label{defJ}
\end{equation}
and it will also be independent of $\bx$. Making the Fourier expansion
\begin{equation}
{\bf E}^a (t) = \int \frac{d \omega}{2\pi} {\bf E}^a ( \omega ) e^{-i \omega t},
\end{equation}
and similarly for ${\bf J}^a$, we can define the dynamical conductivity,
$\sigma ( \omega )$
by the expected linear response relation
\begin{equation}
{\bf J}^a ( \omega ) = \sigma ( \omega ) {\bf E}^a ( \omega )
\end{equation}

For completeness, let us also note here the expression for the total momentum~\cite{iz},
${\bf P}$
\begin{equation}
{\bf P} = \left\langle \pi_{\alpha} \vec{\nabla}_x \phi_{\alpha} \right \rangle.
\end{equation}
Notice that it is quite distinct from ${\bf J}^a$. In particular, in the absence of
an external potential, ${\bf P}$ is conserved ({\em i.e.} it obeys an equation
of the form $\partial_t {\bf P} + \vec{\nabla} \cdot 
\stackrel{\leftrightarrow}{\bf T}=0$ for some $\stackrel{\leftrightarrow}{\bf T}$),
while ${\bf J}^a$ is not.

We now make the mode expansion
\begin{eqnarray}
\phi_{\alpha} (\bx,t) &=& \int \frac{d^d k}{(2 \pi)^d}
\frac{1}{\sqrt{2 \varepsilon_k}}
\left( a_{\alpha} (\bk, t) e^{i \bk \cdot \bx} + a_{\alpha}^{\dagger} (\bk, t) e^{- i
\bk \cdot \bx} \right) \nonumber \\
\pi_{\alpha} (\bx,t) &=& -i \int \frac{d^d k}{(2 \pi)^d}
\sqrt{\frac{\varepsilon_k}{2}}
\left( a_{\alpha} (\bk, t) e^{i \bk \cdot \bx} - a_{\alpha}^{\dagger} (\bk, t) e^{- i
\bk \cdot \bx} \right),
\label{defpi}
\end{eqnarray}
where the $a(\bk, t)$ operators satisfy the equal-time commutation relations
\begin{eqnarray}
\left[ a_{\alpha} (\bk, t) , a_{\beta}^{\dagger} (\bk', t) \right] &=& 
\delta_{\alpha \beta} (2 \pi)^d \delta^d (\bk - \bk')
\nonumber \\
\left[ a_{\alpha} (\bk, t) , a_{\beta}  (\bk', t) \right] &=& 0. 
\end{eqnarray}
It can now be verified that (\ref{qt1}) is satisfied, and ${\cal H}_0$ is given by
\begin{equation}
{\cal H}_0 = \int \frac{d^d k}{(2 \pi)^d} \varepsilon_k \left[ a_{\alpha}^{\dagger} (\bk ,t)
a_{\alpha} (\bk ,t) + 1/2 \right]
\end{equation}
We will also need the expression for the current ${\bf J}^a$ in terms of 
the $a$ and $a^{\dagger}$. We will only be interested in the case where
the system carries a position-independent current: for this case,
inserting (\ref{defpi}) into (\ref{defJ}), we find
\begin{eqnarray}
{\bf J}^a (t) &=& {\bf J}^a_{I} (t) + {\bf J}^a_{II} (t)
\nonumber \\
{\bf J}^a_{I} (t) &=&  i Q L^a_{\alpha\beta} \int\frac{d^d k}{(2\pi)^d}
\frac{\bk}{\varepsilon_k} \left\langle a^{\dagger}_{\alpha} ( \bk , t) a_{\beta} ( \bk , t) 
\right\rangle
\nonumber \\
{\bf J}^a_{II} (t) &=& - i Q L^a_{\alpha\beta} \int\frac{d^d k}{(2\pi)^d}
\frac{\bk}{2 \varepsilon_k} \left\langle a^{\dagger}_{\alpha} ( - \bk , t) 
a^{\dagger}_{\beta} ( \bk , t) \right\rangle + {\rm H.c.}
\end{eqnarray}
It should be evident that processes contributing to ${\bf J}^a_{II}$
require a minimum frequency of 
$2 m$, and so ${\bf J}^a_{II}$
only contributes to $\sigma_{II} ( \omega )$. We will therefore
drop the ${\bf J}^{a}_{II}$ contribution below and approximate
${\bf J}^a \approx {\bf J}^a_{I}$. The ease with which the high frequency components
of $\sigma ( \omega )$ can be separated out is an important advantage of the present
formulation of the quantum transport equations.

A central object in transport theory is the Green's function~\cite{kb,pd}
\begin{equation}
g^{<}_{\alpha\beta} (\bk, \Omega, {\bf R}, t) =
\int \frac{d^d K}{(2 \pi)^d} \int dt_1  
e^{ i {\bf K} \cdot {\bf R} + i \Omega t_1}
\left\langle
a^{\dagger}_{\beta} ( \bk - {\bf K}/2 , t - t_1 /2) a_{\alpha} ( \bk 
+ {\bf K}/2, t + t_1 /2 ) \right\rangle
\label{defg}
\end{equation}
For the case of a system carrying a spatially independent current, 
$g^{<}$ will be independent of ${\bf R}$ and this will implicitly
be assumed below by dropping the ${\bf R}$ argument. 
We also define the particle distribution function
\begin{equation}
f_{\alpha\beta} ( \bk , t) = \int \frac{d \Omega}{2 \pi}
g^{<}_{\alpha\beta} ( \bk, \Omega, t),
\end{equation} 
in terms of which the current is
\begin{equation}
{\bf J}^a (t) =  i Q L^a_{\alpha\beta} \int \frac{d^d k}{(2 \pi)^d}
\frac{\bk}{\varepsilon_k} f_{\alpha\beta} ( \bk, t)
\label{cr1}
\end{equation}
The corresponding expression for the momentum density is
\begin{equation}
{\bf P} (t) =  \int \frac{d^d k}{(2 \pi)^d}
\bk f_{\alpha\alpha} ( \bk, t).
\label{cr2}
\end{equation}
Notice the difference in the structure of the $O(n)$ indices 
between (\ref{cr1}) and (\ref{cr2}).

We note here that our formulation of the transport theory in terms of Green's functions
of the $a_{\alpha}$, $a_{\alpha}^{\dagger}$ rather than those of the $\phi_{\alpha}$,
$\pi_{\alpha}$ was motivated in part by recent exact results in $d=1$ for time-dependent
correlations of models equivalent to ${\cal S}$ for $n=1,3$~\cite{apy,o3}.
In the latter cases it was evident that the physics is  most simply described by
following the propagators of particles created/annihilated 
by $a^{\dagger}_{\alpha}$/$a_{\alpha}$
through their collisions.

\subsection{Collisionless transport}
\label{clesstrans}

In this section we will examine the transport equations for the collisionless case
where ${\cal H}_{\rm int} = 0 $; we 
remind the reader that interactions have already been included at the Hartree-Fock 
level in ${\cal H}_0$ (see Appendix~\ref{susc}). 
Strictly speaking, we also have to remember that the mass $m$ 
can in general depend upon ${\bf E}^a$; however for the case of a momentum-independent
local interaction $u_0$, such a `vertex' correction vanishes to lowest order in ${\bf E}^a$,
and will therefore be omitted from our discussion.

While it is possible to discuss the general $O(n)$ case,
in the interest of simplicity and to keep the physical content transparent,
we will restrict our attention here to the special case $n=2$. The generalization
to $n>2$ is discussed in Appendix~\ref{appo3}.

For $n=2$ ($\alpha = 1,2$), there is only one real antisymmetric matrix, 
and therefore the index $a$
can be dropped. We choose $L_{1,2} = - L_{2,1} = 1$ and $L_{1,1} = L_{2,2} = 0$.
This matrix is off-diagonal and it is helpful to transform to a basis where the
external field is diagonal. We therefore define
\begin{equation}
a_{\pm} (\bk , t) \equiv \frac{a_1 (\bk t) \pm i a_2 (\bk , t)}{\sqrt{2}}
\end{equation}
The current now becomes
\begin{eqnarray}
{\bf J} &=&  Q \int\frac{d^d k}{(2\pi)^d} \sum_{\lambda}
\lambda \frac{\bk}{\varepsilon_k} 
\left\langle a^{\dagger}_{\lambda} ( \bk , t) a_{\lambda} ( \bk , t) \right\rangle
\nonumber \\
&=&  Q \int\frac{d^d k}{(2\pi)^d} \sum_{\lambda}
\lambda \frac{\bk}{\varepsilon_k} 
f_{\lambda} ( \bk , t)
\label{cless0}
\end{eqnarray}
where the index $\lambda$ is assumed here and below to extend over the values 
$\pm 1$, and $f_{\lambda}$ are the particle distribution functions which are now
diagonal in $\lambda$ space. 
Notice that there are two species of charged particles with charges $\pm Q$:
these are the particle-like and hole-like excitations of the bosonic insulator.
Let us also note the expression for the momentum density
\begin{equation}
{\bf P} = \int\frac{d^d k}{(2\pi)^d} \sum_{\lambda}
\bk 
f_{\lambda} ( \bk , t)
\label{cless0a}
\end{equation}
An important difference between (\ref{cless0}) and (\ref{cless0a}) is 
the $\lambda$ inside the summation in (\ref{cless0}) which is absent from (\ref{cless0a}).
Thus the `charge' current is proportional to the  difference of the particle and hole
number currents, while the momentum density is proportional to their sum.

It is now easy to use standard methods~\cite{kb,pd} to derive the following 
transport equation in the collisionless limit described earlier
\begin{equation}
\left( \frac{\partial}{\partial t} +  \lambda Q {\bf E} (t)  \cdot 
\frac{\partial}{\partial \bk} \right)
 f_{\lambda} ( \bk, t) = 0.
\label{cless1}
\end{equation}
In deriving this equation we have made approximations to the charge density
appearing in ${\cal H}_{\rm ext}$ similar
to those made for ${\bf J}$: upon expressing ${\cal H}_{\rm ext}$ in terms of
the $a$, $a^{\dagger}$ we have dropped all terms involving the product of
two $a$'s or $a^{\dagger}$'s as these will only contribute to the high frequency
$\sigma_{II}$. The equations (\ref{cless0},\ref{cless1}) are therefore accurate
to first order in $u_0$ provided we are limiting ourselves to frequencies $\omega \ll 2m$.

Let us now solve (\ref{cless0},\ref{cless1}) in linear response. In the absence 
of ${\bf E}$, the distribution function has the equilibrium value given by 
Bose function
$f_{\lambda} (\bk, t) = n ( \varepsilon_k )$. We Fourier transform from time, $t$, to frequency 
$\omega$, and parametrize to linear order in ${\bf E}$:
\begin{equation}
f_{\lambda} ( \bk , \omega ) = 2 \pi \delta (\omega ) 
n(\varepsilon_k ) +  \lambda Q \bk \cdot {\bf E} (\omega)
 \psi ( k, \omega ),
\label{cless2}
\end{equation}
where we have used the fact that only ${\bf E}$ breaks spatial rotation invariance
and $O(2)$ symmetry to conclude that $\psi$ is independent of $\bk / k$ and $\lambda$.
Now inserting in (\ref{cless1}), and using $\partial \varepsilon_k /\partial \bk
= \bk / \varepsilon_k$ it is simple to solve for $\psi$ 
to leading order in ${\bf E}$:
\begin{equation}
\psi (k, \omega ) = \frac{1}{-i \omega} \frac{1}{\varepsilon_k}
\left( - \frac{\partial n ( \varepsilon_k )}{\partial
\varepsilon_k} \right)
\end{equation}
Finally we insert in (\ref{cless0}) and deduce the conductivity
\begin{equation}
\sigma ( \omega ) = \frac{2 Q^2 }{-i \omega}
\int \frac{d^d k}{(2 \pi)^d}
\frac{k_x^2}{\varepsilon_k^2} \left( - \frac{\partial n ( \varepsilon_k )}{\partial
\varepsilon_k} \right)
\end{equation}
The real part of this agrees with (\ref{k4}). Notice that the leading factor
of 2 comes from the sum over $\lambda$. The current is therefore carried equally by
the thermally excited particles and holes: they move in opposite directions
to create a state with vanishing momentum but non-zero charge current.

We will see in the next section that this charge current can be relaxed by
collisions among the particles and holes. There is no need to invoke umklapp
scattering as a momentum sink/source (as was done by Cha~{\em et al.}~\cite{cha}) as the
current carrying state does not have a net momentum to begin with.

\subsection{Collision-dominated transport}
\label{sec:ctrans}

We have seen in Section~\ref{clesstrans} that to order $u_0$ ($\epsilon$) the transport
is described by the ballistic motion of undamped particles of two charges.
We now consider the collisions of these particles which appear at order $\epsilon^2$.
As we noted in the discussion below (\ref{k6}) the typical energy of a particle
contributing to the transport was of order $T$. Their collisions will lead to 
a broadening of the single quasiparticle pole of order $\epsilon^2 T$~\cite{s};
for $\epsilon$ small, this broadening is weak, and it is then permissible
to argue in terms of a quasiparticle interpretation. 

Applying the standard methods of transport theory discussed in Refs~\cite{kb,pd},
we generalize (\ref{cless1}) to include the collision term which appears at
order $u_0^2$, and obtain to linear order in ${\bf E}$:
\begin{eqnarray}
\left( \frac{\partial}{\partial t} +  \lambda Q {\bf E} \cdot \frac{\partial}{\partial
\bk} \right)
&& f_{\lambda} ( \bk, t) = - \frac{2 u_0^2}{9} \int \frac{d^d k_1}{(2 \pi)^d} 
\frac{d^d k_2}{(2 \pi)^d} \frac{d^d k_3}{(2 \pi)^d} 
\frac{1}{16 \varepsilon_{k} \varepsilon_{k_1}\varepsilon_{k_2}\varepsilon_{k_3}}
 \nonumber \\
&& \times (2 \pi)^d \delta( \bk + \bk_1 - \bk_2 - \bk_3 ) 
 2 \pi \delta ( \varepsilon_{k}
+ \varepsilon_{k_1} - \varepsilon_{k_2} - \varepsilon_{k_3})
\Bigl\{ \nonumber \\
&&~~~~~~~~~~~~2 f_{\lambda} ( \bk , t) f_{-\lambda} (\bk_1 , t) 
[ 1 + f_{\lambda} ( \bk_2 , t)] [ 1 + f_{-\lambda} ( \bk_3 , t)] \nonumber \\
&&~~~~~~~~~~~~+  f_{\lambda} ( \bk , t) f_{\lambda} (\bk_1 , t) 
[ 1 + f_{\lambda} ( \bk_2 , t)][ 1 + f_{\lambda} ( \bk_3 , t)] \nonumber \\
&&~~~~~~~~~~~~- 2 [1 + f_{\lambda} ( \bk , t)][1 + f_{-\lambda} (\bk_1 , t)]
f_{\lambda} ( \bk_2 , t) f_{-\lambda} ( \bk_3 , t) \nonumber \\
&&~~~~~~~~~~~~- [1 + f_{\lambda} ( \bk , t)][1 + f_{\lambda} (\bk_1 , t)]
f_{\lambda} ( \bk_2 , t) f_{\lambda} ( \bk_3 , t)
\Bigr\}.
\label{trans1}
\end{eqnarray}
The collision term on the right-hand-side of (\ref{trans1}) can also be
obtained by a simple argument based on Fermi's golden rule of the type
described in Ref.~\cite{kb}.
A number of simplifications have been made in deriving (\ref{trans1}), and
we now describe and justify them:
\begin{itemize}
\item
The collision term is initially expressed in terms of 
full two-point Green's functions like $g_{\alpha\beta}^{<}$. However, as is 
conventional~\cite{kb,pd}, we assume
we can neglect damping in these Green's functions and express them in terms
of the particle distribution functions $f_{\alpha\beta}$. This is permissible
for a system with well-defined quasiparticles, as is the case here for small $\epsilon$.
This approximation will not affect the conductivity at order $\epsilon^2$.
\item
In addition to the collision terms, there are also self-energy terms affecting
the quasiparticle energies at order $u_0^2$. These include terms that couple
the usual Green's functions to the anomalous ones (involving expectation values
of pairs of creation or annihilation operators).
Such terms will, in general, modify the
conductivity at order $\epsilon^2$. However, they do not affect the nature of broadening
of the delta function in $\sigma'_I ( \omega )$. The total spectral weight in
$\sigma'_I ( \omega)$ will change from that in (\ref{k6}) at order $\epsilon^2$ due to
these terms, but the functional form of of $\sigma'_I ( \omega )$
at frequencies of order $\epsilon^2 T$ will not be affected.
\item
We have neglected collisions which involve creation or annihilation of particle-hole
pairs as they have negligible phase space. Thus a collision in which {\em e.g.}
a positively charged particle of momentum ${\bk}$ turns into two positively charged
particles and a negatively charged hole with momenta $\bk_1$, $\bk_2$, and $\bk_3$
respectively is permitted by the symmetries of the problem. However, it remains
to evaluate the phase space over which such collisions conserve total energy
and momentum. Notice that the `mass' $m$ of the particles/holes is of order
$\sqrt{\epsilon} T$ (Eqn (\ref{k1a})) while their momentum is of order $T$. So to
leading order in $\epsilon$ we may just replace the energy momentum relation (\ref{k5})
by (see also the discussion below (\ref{k6}))
\begin{equation}
\varepsilon_k = k.
\end{equation}
We will use this simplified energy-momentum relation throughout this subsection.
The particle-hole pair-creation collision now requires that $\bk = \bk_1 + \bk_2
+\bk_3$ and $k = k_1 + k_2 + k_3$. This is only possible if all three momenta
are collinear, and this process therefore has vanishing phase space. More generally,
for a non-zero $m$, the phase space vanishes as $\epsilon \rightarrow 0$.
\end{itemize}

We now insert the parametrization (\ref{cless2}) in (\ref{trans1}), linearize the resulting
equation in the external electric field ${\bf E}$, and obtain a linear integral
equation for the unknown function $\psi ( k , \omega )$. Further, to 
leading order in $\epsilon$,
we may set $d=3$ in the collision term in (\ref{trans1}), and replace the coupling
$u_0$ by its fixed point value for $n=2$ (see Appendix~\ref{susc})
\begin{equation}
u_0 = \frac{24 \pi^2 \epsilon}{5}.
\end{equation}
Further details of this linearization procedure are given in Appendix~\ref{sec:collision}.
The final integral equation for $\psi (k, \omega )$ can be written as
\begin{equation}
- i \omega \psi ( k, \omega ) + \frac{1}{k} \frac{\partial n(k)}{\partial k}
= - \epsilon^2 \int_0^{\infty} dk_1 \left[ F_1 ( k, k_1) \psi(k, \omega) 
+ F_2 (k, k_1 ) \psi ( k_1, \omega) \right].
\label{inteq}
\end{equation}
The expressions for the functions $F_{1,2}$ are quite lengthy and are discussed
in Appendix~\ref{sec:collision}. 

It is now useful to scale out the dependence of all functions on $\epsilon$ and $T$
so that the final integral equation is written in dimensionless form, and has
all couplings of order unity. From the expressions in Appendix~\ref{sec:collision}
we know that $F_{1,2}$ are homogeneous functions of momenta and $T$ which satisfy
\begin{equation}
F_i ( k, k_1) =  \Phi_i ( k/T, k_1 / T)
\end{equation}
for some $\Phi_i$, with $i=1,2$. 
(Here, and everywhere in this section, the dimensional analysis is
performed for at $d=3$: it is not difficult to extend it to general $d$, but we will not
in the interests of simplicity, and because it is unnecessary to obtain results
to leading order in $\epsilon$.) Now by examining (\ref{inteq}) we see that it is useful
to introduce the function $\Psi$ defined by
\begin{equation}
\psi(k, \omega ) = \frac{1}{\epsilon^2 T^3} \Psi \left( \frac{k}{T} , \frac{\omega}{\epsilon^2 T}
\right)
\label{scalepsi}
\end{equation}
In terms of $\Phi_i$, $\Psi$ the integral equation (\ref{inteq}) takes the dimensionless
form
\begin{equation}
- i \widetilde{\omega} \Psi ( \overline{k}, \widetilde{\omega} ) -
\frac{1}{4 \overline{k} \sinh^2 ( \overline{k}/2 ) }
= - \int_0^{\infty} d\overline{k}_1 \left[ \Phi_1 ( \overline{k}, \overline{k}_1) 
\Psi(\overline{k}, \widetilde{\omega}) 
+ \Phi_2 (\overline{k}, \overline{k}_1 ) \Psi ( \overline{k}_1, \omega) \right],
\label{inteq1}
\end{equation}
where $\overline{k} = k/T$ and $\widetilde{\omega} = \overline{\omega}/\epsilon^2
= \omega / \epsilon^2 T$.
Now, from (\ref{cless0}) we see that the conductivity $\sigma_I ( \omega )$ obeys
\begin{equation}
\sigma_I ( \omega ) = \frac{Q^2 T^{d-2}}{\epsilon^2} \Sigma_I \left(
\frac{\omega}{\epsilon^2 T} \right)
\label{inteq3}
\end{equation}
where the scaling function $\Sigma_I$ is given by
\begin{equation}
\Sigma_I ( \widetilde{\omega} ) = \frac{1}{3 \pi^2} \int_0^{\infty} 
\overline{k}^3 d \overline{k} \Psi ( \overline{k} , \widetilde{\omega} ) 
\label{sigmaI}
\end{equation}
We can now already see from the structure of 
(\ref{inteq3}) that $\sigma'_I ( \omega = 0 )$ has a value
of order $T^{d-2}/\epsilon^2$, and that the width of the `Drude' peak
will be of order $\omega \sim \epsilon^2 T$. Further, the collision term will not 
modify the total spectral weight in $\sigma_I$, which will therefore be identical
to that in (\ref{k6}); this implies that the function $\Sigma_I$, as defined by
(\ref{inteq1}) and (\ref{sigmaI}) should satisfy
\begin{equation}
\int_0^{\infty} d \widetilde{\omega} \Sigma'_I ( \widetilde{\omega} ) = \frac{\pi}{18}.
\label{inteq4}
\end{equation}
It should be noted that this sum rule is special to the leading order in 
$\epsilon$ being considered here. For $\epsilon$ of order unity, there is no
sharp distinction between $\sigma_I$ and $\sigma_{II}$ and there is no sum rule: indeed
the integral in (\ref{inteq4}) when carried out over the total $\sigma$ will be 
divergent. For any realistic lattice model there is a large microscopic energy scale
($\sim U$ the repulsion between bosons on the same site) beyond which the universal
scaling results do not apply, and the entire spectral weight (including frequencies
beyond $U$) is not divergent; this latter spectral weight satisfies a sum rule
related to non-universal microscopic quantities, and is unrelated to the universal
result (\ref{inteq4}).

\begin{figure}
\epsfxsize=5.5in
\centerline{\epsffile{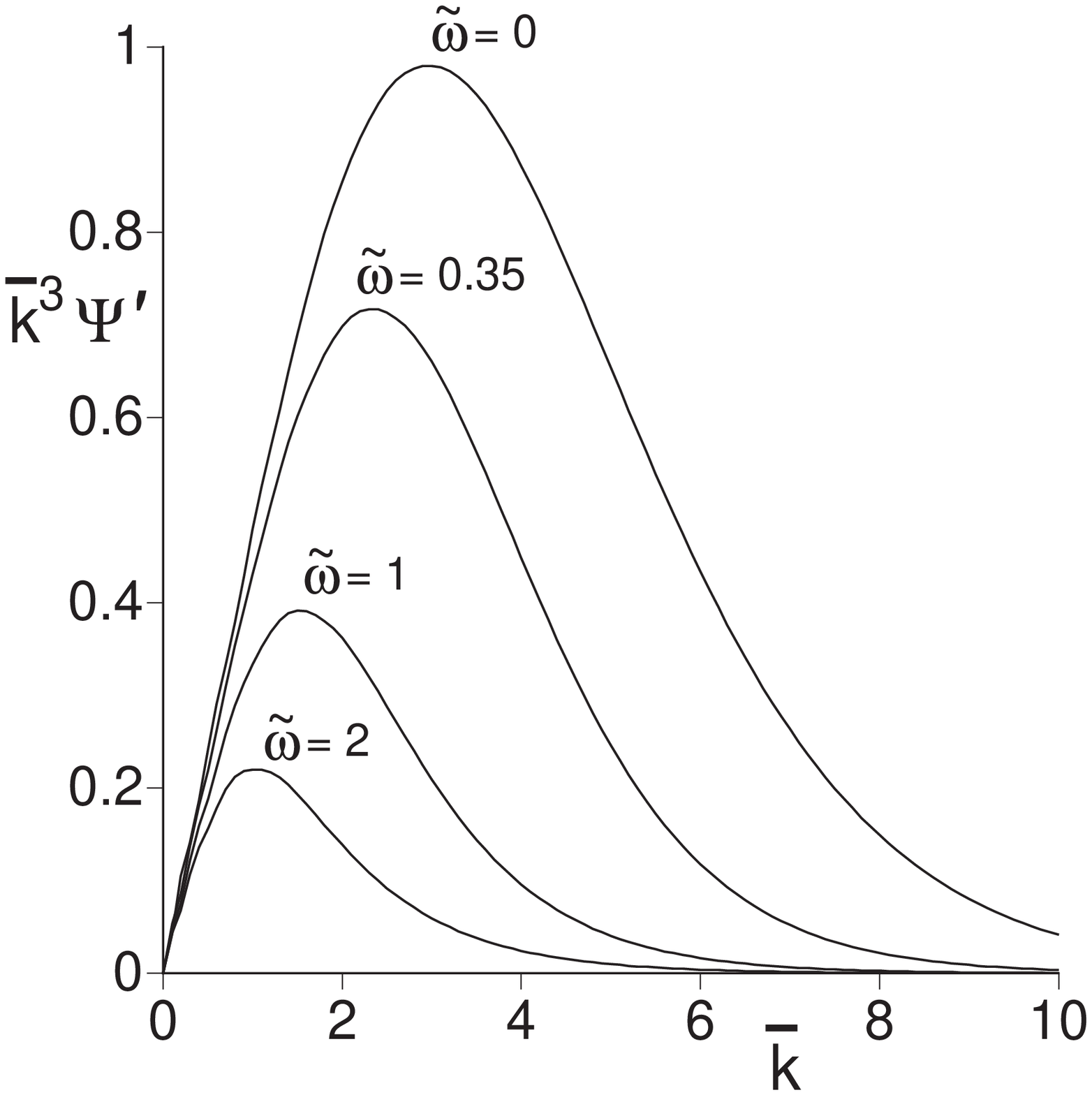}}
\vspace{0.2in}
\caption{
Real part of the universal function $\overline{k}^3 \Psi ( \overline{k}, 
\widetilde{\omega})$ as a function of $\overline{k}$ for a few values
of $\widetilde{\omega}$. The function $\Psi$ is
defined in (\protect\ref{scalepsi}) and (\protect\ref{cless2}), and was
obtained by numerical solution
of the linearized quantum Boltzmann equation (\protect\ref{inteq1}).
At $\widetilde{\omega} = 0$, $\Psi$ is real, but is complex for 
general $\widetilde{\omega}$.
Here $\overline{k} = k/T$,
and $\widetilde{\omega} = \overline{\omega}/\epsilon^2 = \omega/\epsilon^2 T$
(in physical units $\overline{k} = \hbar c k / k_B T$,
$\widetilde{\omega} = \hbar \omega/ \epsilon^2 k_B T$).
}
\label{fig5}
\end{figure}

\begin{figure}
\epsfxsize=5.5in
\centerline{\epsffile{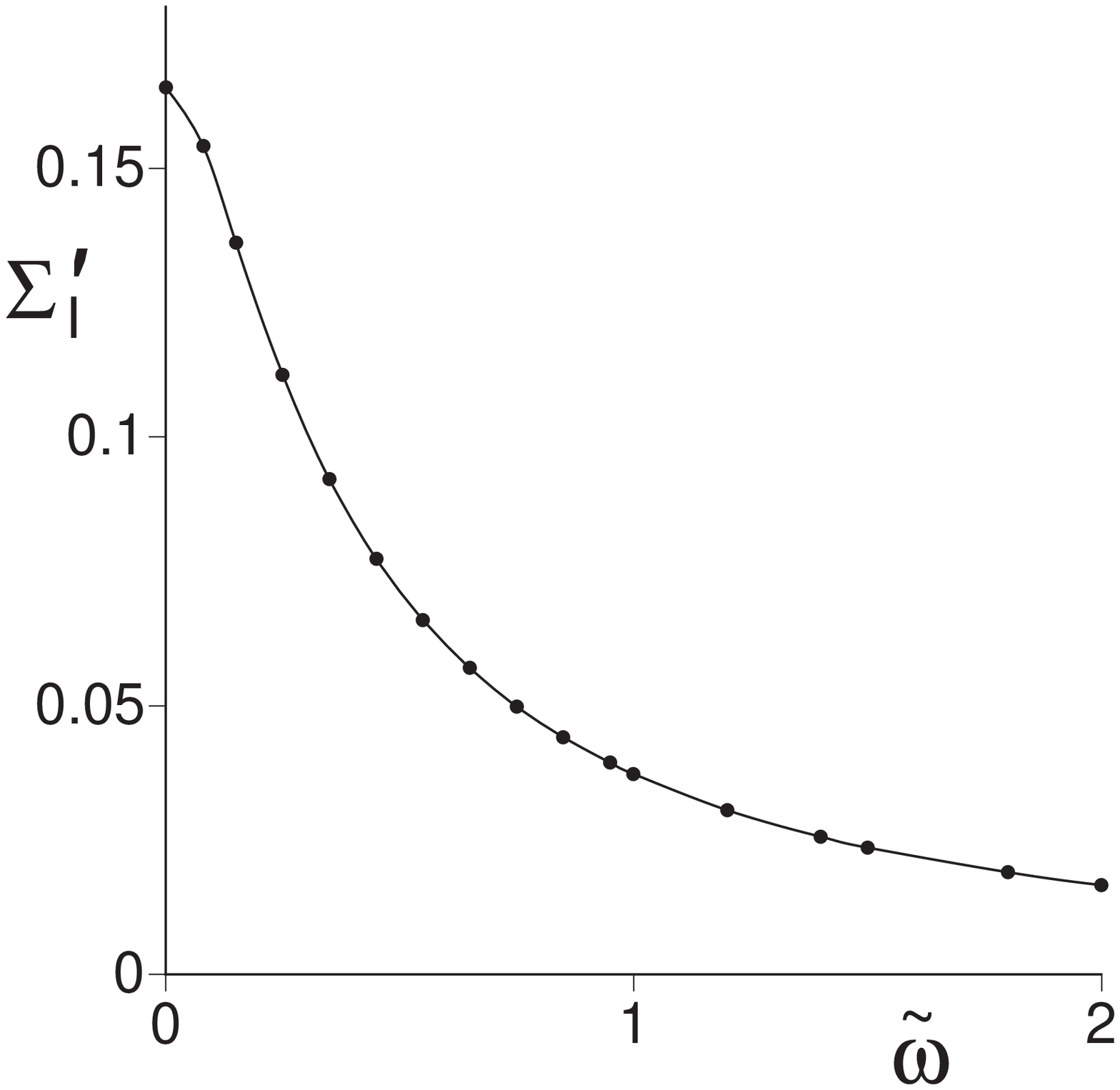}}
\vspace{0.2in}
\caption{
The real part of the universal function $\Sigma_I ( \widetilde{\omega} )$,
which is related to the low frequency part of the
conductivity ($\sigma_I ( \omega ) $) by (\protect\ref{inteq3}).
The results are obtained by the numerical
solution of (\protect\ref{inteq1}), followed by the integration
in (\protect\ref{sigmaI}). This function describes the inelastic
collision-induced broadening
of the $\omega=0$ delta function in Fig~\protect\ref{fig4} at a frequency
scale of order $\epsilon^2 T$. The conductivity has an additional 
continuum contribution ($\sigma_{II} ( \omega )$) at frequencies larger than
$\omega \sim \epsilon^{1/2} T$ which is not shown above (see Fig~\protect\ref{fig3}).
}
\label{fig6}
\end{figure}

It now remains to numerically solve (\ref{inteq1}) to determine $\Psi (\overline{k},
\widetilde{\omega} ) $, and then to obtain $\Sigma_I$ from (\ref{sigmaI}).
The integral equation (\ref{inteq1})
was solved by a straightforward numerical iteration, and we
found very rapid convergence to its unique solution. We show a plot of its
solution at a few values of $\widetilde{\omega}$ in Fig~\ref{fig5}. 
The final result for
the universal function $\Sigma'_1 ( \widetilde{\omega})$ is shown in Fig~\ref{fig6}.

\section{Conclusions}
\label{conc}
The central message of this paper is simply stated: understanding the universal
d.c. conductivity of a two-dimensional system at its quantum critical point
requires a non-zero temperature analysis of the hydrodynamic, collision-dominated
regime where $\hbar \omega \ll k_B T$. 
{\em The transport in this low frequency regime is incoherent, but nevertheless the 
remarkable fact is that the conductivity is still a universal number times $e^2 / h$.}
Computations carried out exactly
at $T=0$, with $\omega \rightarrow 0$,
do not yield the d.c. conductivity, and are controlled by very
different physical processes involving phase-coherent, collisionless transport.
A related comment is that a theoretical analysis for the d.c. conductivity 
must necessarily be formulated in real time, as the imaginary time Matsubara
frequencies are of order $2 \pi k_B T/\hbar$ or larger, and cannot easily capture
the singular structure in Fig~\ref{fig2}.
These criticisms apply not only to computations
of the universal conductivity at the 
superfluid-insulator transition~\cite{fgg,cha,mpaf,wz,kz,fz,ih2,runge,swgy,mtu,bls,wsgy} but
also to the transitions between the Hall plateaus~\cite{ww,cfw}, all of whom computed the
analog of $\Sigma ( \infty )$.
There are also mappings between these models~\cite{klz}, but they presumably hold for
all values of the ratio $\hbar \omega/k_B T$. 
There are also special self-dual models~\cite{fk,lp}
in which the conductivity is claimed~\cite{steve} to be 
independent of $\hbar \omega/k_B T$,
but this is not expected to be the generic physical situation.
The computation of the finite $T$ conductivity at the superfluid-insulator transition
in one dimension by Giamarchi and Schulz~\cite{gs} (and close to one dimension
by Herbut~\cite{ih}) seems closer in spirit to 
our approach, but it would be useful to have an explicit computation as
a function of $\hbar\omega/k_B T$ to verify this.

Many experiments on two-dimensional films appear not to observe
a universal conductivity at the superfluid-insulator transition.
Our results imply that this is very likely due to crossovers caused by
{\em inelastic\/} scattering
mechanisms other than those contained in the critical theory.
Measurements of the conductivity as a function of frequency (Section~\ref{ac})
should help disentangle these effects.

In the remainder of this section we comment on some general experimental and 
theoretical issues related to the results of the paper.

\subsection{Imaginary time Monte Carlo}

Most existing numerical studies~\cite{cha,swgy,mtu,bls,wsgy} 
of $\sigma$ at a two-dimensional
quantum critical point used a Monte Carlo simulation in imaginary time. 
The simulation measures the values of $\sigma$ at the non-zero
Matsubara frequencies {\em i.e.} at $\sigma ( 2 \pi n  T i)$ where $n$
is a non-zero, positive integer. 
The limit $\Sigma (z,  |z| \rightarrow \infty )$ is
expected to be the same number $\Sigma ( \infty )$ for all values of the
phase of $z$, and therefore its value can be deduced from a 
relatively straightforward analysis of the numerical data.
The value of $\Sigma ( 0 )$ is more problematical. All of the interesting
structure in $\Sigma (z)$ discussed in this paper occurs for $z$ of order, or
less than unity, and it is difficult to extract this information from its
values at the
nonzero quantized Matsubara frequencies. 
Cha~{\em et al.}~\cite{cha} numerically examined the model ${\cal S}$
and found little dependence on $\omega_n /T$. Note that this
is the model for which we have computed the strongly $\omega/T$ dependent 
conductivity here, but, as expected, this is apparently not evident at the
imaginary Matsubara frequencies. 
Similarly, Wallin~{\em et al.}~\cite{wsgy}, also found little evidence for
a significant dependence of the critical conductivity of disordered models
on $\omega_n / T$.

We are of the opinion that it will be difficult
to determine $\Sigma (0)$ from this method, unless highly accurate numerical
results are obtained at the imaginary frequencies.
The difficulty is also apparent by a glance at Fig~\ref{fig2}: at $T=0$,
the value of $\Sigma (0)$ appears only at a single point which carries
zero weight under any integral over frequencies.
The accurate numerical data should then be analyzed in the following manner.
First, from observations at a number of different values of $\omega_n$ and $T$,
the universal scaling part of $\sigma$, dependent only on the ratio $\omega_n / T$,
should be obtained. Note that this universal scaling part reaches a constant
value ($(4 e^2 /\hbar) \Sigma (\infty)$) as $\omega_n \rightarrow \infty$,
while the full $\sigma$ vanishes as $1 / \omega_n^2$ for frequencies larger than
a microscopic lattice scale. Then a Pade analysis, general enough to allow the
structure in Fig~\ref{fig1}, should be used to analytically continue {\em only\/} the
universal scaling part to real frequencies.

It would also be interesting to explore 
newer methods: perhaps examining an open system in
which it is possible to have a net current in thermodynamic equilibrium,
or doing a computation in real time. The exact diagonalization approach
of Runge~\cite{runge} can perhaps be extended to $T>0$ without a great
deal of difficulty.

\subsection{Dangerously irrelevant interactions}
\label{dii}

It is possible to violate the basic scaling result (\ref{tautr}) if quantum-mechanical 
interactions between the elementary critical excitations happen to be dangerously 
irrelevant~\cite{sro}; in that case we expect
\begin{equation}
\frac{1}{\tau_{\rm tr}} \sim  u T^{1 + \theta_u /z}.
\label{taudi}
\end{equation}
where $u$ is proportional to some power of the irrelevant interaction, $\theta_u > 0$
is the associated crossover exponent, and $z$ is the dynamic critical exponent.
Anderson localization transitions, with interactions leading only to a infrared
cutoff as a phase-breaking rate, are a realization of such a scenario.
(Quantum-impurity critical points~\cite{uw,fls,lss,kf} have
some similarities to bulk systems with dangerously irrelevant interactions, and are
discussed in more detail in Appendix~\ref{dqm}.)  However, this scenario is much
less likely to be realized
 at {\em bulk} two-dimensional quantum critical points, and we consider
it unlikely that interactions can be neglected
 for the superfluid-insulator transition. 
The quantum Hall transition has primarily been studied using non-interacting
electrons~\cite{cc}, but there is
evidence that Coulomb interactions are relevant at quantum Hall transitions~\cite{ps,lw}.
Experimental measurements of $z$ in the quantum Hall system~\cite{wei} 
indicate the value $z=1$ for the dynamic critical exponent, and this is incompatible
with free electron models. The a.c. conductivity measurements of 
Engel {\em et al.}~\cite{engel} in a quantum Hall system
(discussed further in Section~\ref{ac} below) show a characteristic frequency
scale $\omega \sim k_B T / \hbar$ which is inconsistent with the result (\ref{taudi}) 
for the case of irrelevant interactions, but is consistent with the interacting
theory result (\ref{tautr}).

\subsection{Luttinger Liquids}

A well studied critical system in $d=1$ is the
gapless Luttinger liquid ground state of interacting fermions or bosons away
from commensurate filling fractions. Here we discuss how this familiar
system fits into the general framework of this paper. 
The non-zero $T$ conductivity of the critical theory of the Luttinger liquid
is given by
\begin{equation}
\sigma' ( \omega ) = K \delta ( \hbar \omega ),
\label{ll1}
\end{equation}
where $K$ is some $T$ independent constant. Notice that there is no broadening
of the delta function, even for $T>0$, in the scaling limit. (Related
considerations also apply to Fermi liquids in $d>1$~\cite{rs,nw}, but
for many purposes these are better thought of as the analog of Goldstone
phases rather than critical phases.)
Let us now rewrite (\ref{ll1}) as
\begin{equation}
\sigma' ( \omega ) = K (k_B T)^{-1} \delta \left( \frac{\hbar \omega}{k_B T} \right).
\label{ll2}
\end{equation}
Now notice that (\ref{ll2}) is consistent with the general scaling form 
(\ref{scaling}) for $d=1$, and the known dynamic exponent $z=1$.
The scaling function $\Sigma$ takes the simple and singular form
$\Sigma' (\overline{\omega} ) = \delta (\overline{\omega})$. For this special
form of $\Sigma$, the limits $\omega \rightarrow 0$ and $T \rightarrow 0$ do
happen to commute. 

These singular properties of the critical theory of the Luttinger
liquid are clearly a consequence of the absence of scattering between
carriers in the critical theory: like many other critical theories in $d=1$,
it is conformally invariant, and correlation functions (including those for $T>0$)
factorize into independent components given by the left and right movers.
In contrast, for critical theories in $d>1$, like the one studied in this paper,
no such analogous factorization exists. 
Further, for $d>1$ models below their upper critical dimension, there is scattering
between the carriers already in the scaling limit. 
To introduce scattering in the Luttinger liquid,
it is necessary to go beyond the scaling limit, and consider corrections to scaling~\cite{gm}.
There corrections are therefore dangerously irrelevant, and will destroy 
simple delta function form of the conductivity in (\ref{ll1}), and
the limits $\omega \rightarrow 0$ and $T \rightarrow 0$ will no longer commute~\cite{gm}.

\subsection{Measurements of the a.c. conductivity}
\label{ac}

A finite $\omega$ measurement of the conductivity in a situation related
to that discussed in this paper has been carried out by Engel~{\em et al.}~\cite{engel}.
They examined the transition between integer quantum Hall plateaus by studying
the dependence of the conductivity on $\omega$, and the deviation of the field
from its critical value $\delta \equiv (B-B_c)/B_c$. This should obey~\cite{shahar}
the scaling form
\begin{equation}
\sigma = \frac{e^2}{\hbar} \widetilde{\Sigma} \left( \frac{\hbar \omega}{k_B T}, 
\frac{\delta}{T^{1/\nu z}} \right),
\label{scaling2}
\end{equation}
which generalizes (\ref{scaling}) to $\delta \neq 0$. 
In their analysis, Engel~{\em et al.}~\cite{engel} 
focussed mainly on the $\delta$ dependence at
$\hbar \omega \gg k_B T$: they measured the width of the transition region
in $\delta = \Delta B$, and found $\Delta B \sim \omega^{1/\nu z}$, in agreement with 
(\ref{scaling2}). 
Further, the $\omega$ dependence of $\Delta B$ saturated for $\omega < k_B T/\hbar$,
in agreement with the ideas we have discussed here~\cite{s,shahar}.
However, Engel~{\em et al.}~\cite{engel} did not analyze the $\omega $ and 
$T$ dependence of $\sigma$ precisely at the critical field $\delta = 0$. It appears
to us that it should be relatively straightforward to extend their measurements
at $\delta = 0$ to test the validity of the scaling form (\ref{scaling}). 
Further, it should also be possible to theoretically determine the expected
form of the scaling function $\Sigma$ in the models for the quantum Hall transitions
considered in Refs~\onlinecite{ww,cfw}.

A limited test of (\ref{scaling}) should also be possible in
measurements of the a.c. conductivity at relatively low frequencies at which
$\hbar \omega \ll k_B T$. At such frequencies, we can expand the scaling
function $\Sigma ( \overline{\omega} )$ about $\overline{\omega} = 0$, and 
analyticity of $T>0$ properties then implies the form
\begin{equation}
\sigma' ( \omega) = \frac{Q^2}{\hbar} \left( \frac{k_B T}{\hbar c} \right)^{(d-2)/z}
\left[ \Sigma (0) - \frac{1}{2} \Sigma^{(2)} (0) \left( \frac{\hbar \omega}{k_B T} \right)^2
+ \ldots \right],
\label{scaling3}
\end{equation}
where $\Sigma (0)$ and $\Sigma^{(2)} (0)$ are expected to be positive universal
numbers of order unity. 
Experimentally, one can test if the frequency-dependent correction
to $\sigma$ has the $1/T^2$ dependence predicted by (\ref{scaling3}).

There do not appear to be any existing measurements of the a.c. conductivity
near quantum critical points in other systems. The results (\ref{scaling}) and 
(\ref{scaling3}) should apply also to the superfluid-insulator transition in thin 
films~\cite{hebard,lg}, to the quantum transition in the doped cuprates~\cite{gb}
and two-dimensional 
MOSFETs~\cite{sarachik}, and to the metal-insulator transition in 
three-dimensional Si:P~\cite{bk,lohneysen}; we hope that experiments will be undertaken,
as the results will be central to our theoretical understanding of these systems.
In particular, if scaling as a function
of $\hbar \omega/k_B T$ is observed, it would establish quite conclusively that
interactions are an essential ingredient in the critical theory.

Also, we note that our picture suggests a rather interesting
non-monotonic $\omega$ dependence of $\sigma$ in at the metal-insulator transition in
$d=3$: for small $\hbar\omega/k_B T$
$\sigma$ should decrease with increasing $\omega $
as predicted by (\ref{scaling3}), but for larger $\hbar \omega/k_B T$ 
it should increase
as $\sim \omega^{1/z}$. It would be worthwhile to undertake analytic calculations
for disordered, interacting, electronic systems to search for this non-monotonic
$\omega$ dependence: there are cases where the critical theory is accessible
at low orders in the $2+\epsilon$ expansion~\cite{bk}.
There are existing calculations for the $\omega$ and $T$ dependence of the
conductivity in the weakly disordered metal~\cite{aa} in which $1/\tau_{\varphi} 
\ll k_B T/\hbar$; these need to be extended to the critical point
where we expect that $1/\tau_{\varphi} \sim k_B T/\hbar$.

\subsection{Self duality}
\label{duality}
Of great current interest is the issue of ``boson-vortex duality''
at two-dimensional quantum critical points. At the superfluid-insulator transition,
many of the experimentally measured values of $\Sigma (0)$ appear 
to be tantalizingly close~\cite{lg}
to a value predicted by self-duality arguments~\cite{mpaf,wz}. For the quantum Hall
transitions, experimental evidence for self-duality has been presented
recently~\cite{stsss}. This self-duality appears rather surprising as there is
no fundamental reason for an equivalence between the underlying boson and vortex
Hamiltonians~\cite{mpaf}. We suggest here that these inequivalent Hamiltonians
will be apparent in the value of the $T=0$ conductivity, given by $\Sigma ( \infty )$,
which, incidentally, is the quantity that has been explicitly or implicitly studied
 in earlier 
work~\cite{fgg,cha,mpaf,wz,kz,fz,ih2,runge,swgy,mtu,bls,wsgy,ww,cfw,klz,fk,lp}. 
In contrast, the d.c. conductivity, given
by $\Sigma (0)$, is controlled by the crossover to
the hydrodynamic, collision dominated regime, and
we propose that this could 
be less sensitive to the details of the boson-vortex Hamiltonian.
This proposal is easily subject to experimental tests: measure $\Sigma ( \infty )$
by determining the conductivity
in the regime $\hbar \omega \gg k_B T$, and see if the self-duality predictions
continue to hold---we predict they will {\em not}. 

Our suggestion that $\Sigma ( 0 ) $ could be insensitive to differences in
the interaction Hamiltonian of bosons and vortices
is motivated by the expectation that an important role of interactions is in
the collisions that establish local thermodynamic equilibrium~\cite{kb,ruzin}. As a result,
the equations governing the net
hydrodynamic flow of bosons and vortices in the collision dominated regime
could be more symmetrical than the underlying Hamiltonian.
In other words, we are proposing here
that a true understanding of the
experimentally observed 
duality near the quantum critical point
will emerge from a study of the crossover 
from the microscopic quantum-critical physics of the elementary
excitations to the low frequency
collision dominated regime best described by a 
quantum Boltzmann equation.

We suggest that even the simple model
${\cal S}$ is self-dual for $d=2$, $n=2$ in its d.c. transport
{\em i.e.} $2 \pi \Sigma (0) = 1$ exactly for the
two dimensional quantum XY rotor model. 
This model has boson particle/hole excitations with short-range interactions,
while the vortices in the dual representation have long-range logarithmic (Coulomb in
$d=2$)
interactions. However, at non-zero $T$, this logarithmic interaction will be
screened in manner analogous to the 
classical Debye screening above the Kosterlitz-Thouless temperature, $T_{KT}$; indeed the
region $g=g_c$, $T> 0$ of interest here is continuously connected to the
$g<g_c$, $T> T_{KT}$ region.
So the effective interactions between the bosons and vortices are both short-range
for $T>0$, leaving open the possibility of self-dual behavior at low frequencies.
There is now general agreement
that $2 \pi \Sigma ( \infty ) \approx 0.3 $ for this model~\cite{fz}.
This paper contains the first computation of $\Sigma (0)$, and
as we noted earlier near Eqns (\ref{i1},\ref{i1b}), it is quite remarkable,
though possibly fortuitous,
that our leading order result for $\Sigma (0)$ in the 
$\epsilon$ expansion differs from the self-dual value by
less than 4\%. 
Definitively establishing this self-duality would however
require techniques other than expansion in $\epsilon = 3-d$, or $1/n$,
as it is only possible precisely at $d=2$, $n=2$. It would be of great
interest to undertake higher precision quantum Monte Carlo, exact diagonalization,
or high temperature series studies while carefully examining the reliability of 
the analytic continuation to real frequencies.

\acknowledgements
We thank R.N.~Bhatt, S.~Chakravarty, 
M.~Chertkov, S.~Das~Sarma, D.~Shahar, T.~Giamarchi, A.F.~Hebard, 
G.~Kotliar, D.H.~Lee, F.~Lesage, A.~MacDonald, A.~Millis, 
H.~Saleur, R.~Shankar, T.~Senthil, D.~Tsui, S.~Zhang, G.~Zimanyi and especially 
E.~Fradkin, M.P.A.~Fisher, S.M.~Girvin, S.~Kivelson 
and N.~Read
for helpful discussions. In addition, our interest in this subject was stimulated
by early discussions with N.~Trivedi, S.L.~Sondhi asked some probing and
invigorating questions, and E.~Fradkin provided the stimulus for the 
discussion in Appendix~\ref{dqm}.
This research was supported by the National Science Foundation 
Grant DMR-96-23181. 
\newline
\newline
{\em Note added:} The reader may be interested in a recent article~\cite{geilo} 
by one of us
discussing this work in the context of recent results on 
non-zero temperature 
dynamical properties of ${\cal S}$ for other values of $d$ and $n$,
including exact solutions in $d=1$.

\appendix

\section{Transport in dissipative quantum mechanics}
\label{dqm}
We will consider the transport properties of a well-studied and representative
model from the subject of dissipative quantum mechanics
and related quantum-impurity problems~\cite{uw}. 
Our purpose here is to contrast the $\omega$ and $T$ dependence of
the transport coefficient in such a situation, with that of bulk critical points studied in the
body of the paper. We believe that such an exercise will help clarify the significance of our
results for the reader. We thank
E.~Fradkin for posing the questions which led to the analysis below.
We will use units with $\hbar=k_B=1$ in this appendix.

Consider the motion of a quantum `particle' in a periodic
potential in the presence of a linear coupling to a Ohmic heat 
bath~\cite{uw,schmid,ghm,fzwerg,fsw,fles,lesa,kf1,moon,fls,lss,kf}. 
We represent the time-dependent
co-ordinate of the particle by $X (t)$, and its analytic continuation to imaginary
time by $X(\tau)$, and the Fourier transform to imaginary
Matsubara frequencies by $X(\omega_n)$.
The imaginary time
effective action obtained after
integrating out the degrees of freedom of the heat bath is
\begin{equation}
S_1 = \frac{T \alpha}{4\pi} \sum_{\omega_n} |\omega_n| \left|
X ( \omega_n ) \right|^2 - y \int d \tau \cos (X(\tau)).
\label{dqm1}
\end{equation}
Here $\alpha$ is a dimensionless coupling constant characterizing the
strength of the Ohmic dissipation, and $y$ measures the strength of the periodic potential.
Models like (\ref{dqm1}) describe tunneling in a SQUID, where $X$ is interpreted
as the flux in the SQUID, or tunneling between Luttinger liquids and quantum Hall
edge states~\cite{fls,kf,kf1,moon}, where $X$ now becomes a bosonic phase field.

The action (\ref{dqm1})
can be written in a form local in time if we extend the quantum degree of freedom
$X(t)$ to an infinite number of degrees of freedom $X (x,t)$ lying along the line
$-\infty < x < \infty$, with $X(t) \equiv X(x=0,t)$; then $S_1$ is equivalent 
to~\cite{ghm,fsw,fles,kf1,fls,lss,kf}:
\begin{equation}
S_2 = \frac{\alpha c}{8 \pi} \int dx d\tau \left(
(\partial_x X(x,\tau))^2 + \frac{1}{c^2} ( \partial_{\tau} X(x,\tau))^2 \right)
 - y \int d \tau \cos (X(\tau)),
\end{equation}
where $c$ is an arbitrary velocity.
Notice that the cosine interaction acts only along the single line $x=0$,
identifying this as a boundary critical phenomena problem.

We shall be interested here in the linear 
response of the system to a time-dependent force, $F(t)$, acting on the particle.
In imaginary time, in the presence of such a force
\begin{equation}
S_1 \rightarrow S_1 - \int d \tau F(\tau) X(\tau).
\end{equation}
In the presence of a time-independent force, we expect the particle to acquire
a finite velocity, $V \equiv d X/dt$, in steady state: this allows us to define
a {\em mobility}, $G$, by $V = G F$. More generally, we expect a frequency dependent
response $G(\omega)$, defined by
\begin{equation} 
V(\omega ) = G(\omega ) F(\omega).
\end{equation}
It is our purpose here to describe the behavior of the dynamic 
mobility, $G(\omega)$, at low $\omega$
and $T$, and to compare it with the results for $\sigma ( \omega )$ obtained
in the main part of the paper.
We also note that $G(\omega)$ is the {\em conductance\/} associated with tunneling between 
quantum Hall
edge states~\cite{kf} 
at a value $g=1/\alpha$ in the notation of Moon {\em et al.}~\cite{moon}.

Consider first the properties of $S_1$ under a $T=0$ renormalization group 
transformation under which $\tau \rightarrow \tau e^{-\ell}$. It is 
found~\cite{schmid,ghm,fzwerg}
that $\alpha$ remains unrenormalized, while the potential strength $y$
obeys the simple flow equation
\begin{equation}
\frac{dy}{d\ell} = - \left( \frac{1}{\alpha} - 1 \right) y.
\label{dqm3}
\end{equation}
For $\alpha\neq 1$, this flow has fixed points only 
at $y=0$ and $y=\infty$ (we will not consider the case $\alpha=1$ here).
For $\alpha < 1$, the $y=0$ fixed point is stable and the $y=\infty$ fixed
point is unstable; for $\alpha >1$ the stability of the fixed points is interchanged.
Notice that {\em both} fixed points are free field theories, supplemented by free or
fixed boundary conditions at $x=0$. This is a crucial difference from the bulk
theory ${\cal S}$ in which the critical theory is interacting for $d<3$,
and is primarily responsible for the differences in the structure of $G(\omega)$
and $\sigma ( \omega )$ we shall find below.

Let us now write down the general scaling predictions for $G(\omega)$ which 
follow from the renormalization group arguments. The system will be completely
characterized by a single nonuniversal energy scale, $T_K$, which measures its deviation
from the {\em unstable} fixed point. From the flow equations (\ref{dqm3}) we 
can deduce $T_K \sim y^{\alpha/(\alpha-1)}$ for $\alpha > 1$; 
the perturbation theory to be discussed below shows  
$T_K \sim y^{-\alpha/(1-\alpha)}$ also for $\alpha < 1$. The mobility $G$ has a zero scaling
dimension, and therefore obeys the scaling form
\begin{equation} 
G ( \omega ) =  {\cal G}_{\alpha} \left( \frac{ \omega}{ T} , \frac{T}{T_K} \right)
\label{dqm4}
\end{equation}
where ${\cal G}_{\alpha}$ is a universal function.
We now describe the low $T$, $\omega$ form of ${\cal G}_{\alpha}$ for the cases $\alpha <1$
and $\alpha > 1$ separately.

Consider first $\alpha < 1$. In this case $y$ flows to zero, and therefore
the periodic potential has vanishing
strength at long times and the particle is delocalized. Naive perturbation theory in $y$
is expected to be reliable. The mobility can computed in this perturbation theory
using a Kubo formula: such a computation was carried out in Appendix A of
Ref~\cite{kf1}, and we find from their results
\begin{equation}
G( \omega ) =  
\frac{2 \pi}{\alpha} - \left( \frac{T}{T_K} \right)^{2(1/\alpha -1)} {\cal P}_{\alpha} \left(
\frac{\omega}{ T} \right) + \ldots ,
\label{dqm5}
\end{equation}
where the ellipses represent terms which are 
higher order in  $T/T_K$---this result holds for $\omega \ll T_K$, $T \ll T_K$, but
$\omega/ T$ can be arbitrary.
Notice that (\ref{dqm5}) is
consistent with (\ref{dqm4}).
The explicit
form of ${\cal P}_{\alpha}$ can be deduced after some analysis of Eqn (A7) 
of Ref~\cite{kf1}:
\begin{eqnarray}
{\cal P}_{\alpha} ( \overline{\omega} ) = && 
\left| \Gamma \left( \frac{1}{\alpha} + \frac{i \overline{\omega}}{2 \pi}
\right) \right|^2 \frac{\sinh (\overline{\omega}/2)}{\overline{\omega}} \nonumber \\
&&~~~~~~~~~~~~ - i
\frac{\tan ( \pi/\alpha)}{\overline{\omega}} \left\{
\left| \Gamma \left( \frac{1}{\alpha} + \frac{i \overline{\omega}}{2 \pi}
\right) \right|^2 \cosh(\overline{\omega}/2) - \Gamma^2 \left( \frac{1}{\alpha}
\right) \right\} \nonumber \\
{\cal P}_{\alpha} ( \overline{\omega} \rightarrow 0 ) = &&
\frac{1}{2} \Gamma^2 \left( \frac{1}{\alpha} \right) \nonumber \\
{\cal P}_{\alpha} ( \overline{\omega} \rightarrow \infty ) = &&
\left( \frac{\overline{\omega}}{\pi} \right)^{2(1/\alpha - 1)}
2^{1-2/\alpha} \left[ 1 - i \tan(\pi/\alpha) \right] \nonumber \\
{\cal P}_{1/2} ( \overline{\omega} ) = && \frac{1}{2} + \frac{\overline{\omega}^2}{8 \pi^2}~~,
~~\mbox{a simple special case~\protect\cite{guinea,kf1}}.
\label{dqm9}
\end{eqnarray}
The overall normalization of ${\cal P}_{\alpha}$ is arbitrary, as it can be absorbed
into a redefinition of $T_K$.
The most important property of (\ref{dqm5},\ref{dqm9}) is that the leading term in the mobility,
$ 2 \pi/ \alpha$, is independent of $\omega/T$. 
Alternatively stated, we have
\begin{equation}
G ( \omega \rightarrow 0, T=0) = G(\omega=0, T \rightarrow 0) = \frac{2 \pi}{\alpha}.
\end{equation}
However, it is clear that this leading
term is a property of the $y=0$ fixed point, and the independence on $\omega / T$
is a consequence of its free field nature. The next, subdominant,
 term arises from the leading
irrelevant operator in the theory, and does indeed lead to a non-trivial dependence
on $\omega / T$ in the universal function ${\cal P}_{\alpha}$. In contrast, the bulk model
${\cal S}$, not being a free-field at the critical point, 
had such a $\omega /T$ dependence already in the leading term, without the
inclusion of any irrelevant operators.
This distinction is the central point of this appendix.  

Now turn to $\alpha > 1$. 
In this case $y$ flows to $\infty$, implying that the periodic potential localizes
the particle at $T=0$. The low $T$ mobility clearly cannot be computed by a naive
perturbation theory in $y$. Instead one can use a self-duality property of 
$S_1$~\cite{schmid,fzwerg,kf1} under which $\alpha \leftrightarrow 1/\alpha$,
and then use perturbation theory. The $y=\infty$ fixed point now implies that the
leading scaling result for the mobility is simply $G=0$, and so the $\omega \rightarrow 0$
and $T \rightarrow 0$ limits commute again for a trivial reason.
The leading $\omega$ and $T$ dependence is given by an irrelevant operator,
which now yields
\begin{equation}
G ( \omega ) =  \left(\frac{T}{T_K}\right)^{2(\alpha-1)} {\cal P}_{1/\alpha}
\left( \frac{\omega}{ T} \right) + \ldots,
\label{dqm10}
\end{equation}
where the scaling function ${\cal P}_{1/\alpha}$ was defined in (\ref{dqm9}),
the corrections are higher order in $T/T_K$,
and the result holds for $\omega \ll T_K$, $T \ll T_K$,
but $\omega/ T$ arbitrary.
Notice also the result
\begin{equation}
G(\omega \rightarrow 0, T=0) = 
G( \omega =0, T \rightarrow 0) =0,
\end{equation}
which arises from the $y=\infty$ free field fixed point.

We also note for completeness the high $T$ behavior, with
$T \gg T_K$, to allow contact with the quantum Hall 
edge state tunneling results
of Ref~\cite{moon}. The key is to note that the
$\alpha \leftrightarrow 1/\alpha$ duality
interchanges small and large $y$.
So the result (\ref{dqm10}) also describes the $T \gg T_K$ limit of 
the $\alpha < 1$ case, while (\ref{dqm5}) describes $T \gg T_K$
for $\alpha > 1$. The main change is that the values of $T_K$ (denoted
$\overline{T}_K$) inserted
in these expressions will not be the same as those in the low $T$ limit.
Perturbation theory cannot determine the universal relationship between $T_K$
and $\overline{T}_K$: this requires use of the exact integrability 
of $S_1$, as discussed
in Refs~\cite{fls,fsw,fles,lesa}.

Finally, we suggest that 
the above commutativity of the limits $\omega \rightarrow 0 $
and $T \rightarrow 0$ (and also $\omega \rightarrow \infty$, $T\rightarrow \infty$, where
$\infty$ refers to a scale much larger than $T_K$), and the free field nature of the
fixed points may be the reason for the success of the imaginary time Monte Carlo
simulation of Moon et al.~\cite{moon}.

\section{General transport equation}
\label{appo3}

Here, we 
will generalize the transport equation (\ref{trans1}) of the model ${\cal S}$
to arbitrary $n \geq 2$.

First consider $O(n=3)$.
This case applies to quantum antiferromagnets~\cite{csy} and the external `potentials'
$U^a$ ($a = 1,2,3$) correspond to the three components of a magnetic field.
Let us take the field pointing in the 3 direction: $U^a = (0,0, H)$.
For the antisymmetric matrices we choose $L^{a}_{\alpha\beta} = \epsilon_{a\alpha\beta}$,
the third-rank antisymmetric tensor. 
Note that such a field is the same as that in (\ref{vecpot}).
The distribution functions become diagonal
by changing basis from $a_{1,2,3}$ to
\begin{equation}
a_{\pm} (\bk , t) \equiv \frac{a_1 (\bk, t) \pm i a_2 (\bk , t)}{\sqrt{2}} ~,~
 a_3 ( \bk , t)
\end{equation}
The particle distribution function now has the three diagonal components
$f_{\pm}$, $f_3$. The `magnetization' current is non-zero only in the $a=3$ direction
and equals
\begin{equation}
{\bf J}^3 = 
Q \int\frac{d^d k}{(2\pi)^d}  \frac{\bk}{\varepsilon_k} \left[
f_+ ( \bk , t) - f_- ( \bk , t) \right]
\end{equation}
A key simplifying feature is that the distribution function $f_3$ must
be even in the external field, and will therefore only be modified at quadratic
order in $H$. We will be satisfied by working in linear response, in which
case $f_3 ( \bk , t ) = n ( \varepsilon_k )$.

The $n>3$ case is very similar: the only difference is that there are
now $n-2$ values of $\alpha$ for which $f_{\alpha} ( \bk , t) = n (\varepsilon_k )$
in linear response.

The generalization of the transport equation (\ref{trans1}) to $O(n)$ is
now easily obtained by an application of Fermi's golden rule:
\begin{eqnarray}
\left( \frac{\partial}{\partial t} + \lambda Q \vec{\nabla} H \cdot
\frac{\partial}{\partial \bk} \right)
&& f_{\lambda} ( \bk, t) = - \frac{u_0^2}{9} \int \frac{d^d k_1}{(2 \pi)^d} 
\frac{d^d k_2}{(2 \pi)^d} \frac{d^d k_3}{(2 \pi)^d} 
\frac{1}{16 \varepsilon_{k} \varepsilon_{k_1}\varepsilon_{k_2}\varepsilon_{k_3}}
 \nonumber \\
&& \times (2 \pi)^d \delta( \bk + \bk_1 - \bk_2 - \bk_3 ) 
 2 \pi \delta ( \varepsilon_{k}
+ \varepsilon_{k_1} - \varepsilon_{k_2} - \varepsilon_{k_3})
\Bigl\{ \nonumber \\
&&~~~~~~~~~~~~4 f_{\lambda} ( \bk , t) f_{-\lambda} (\bk_1 , t) 
[ 1 + f_{\lambda} ( \bk_2 , t)] [ 1 + f_{-\lambda} ( \bk_3 , t)] \nonumber \\
&&~~~~~~~~~~~~+ 2 f_{\lambda} ( \bk , t) f_{\lambda} (\bk_1 , t) 
[ 1 + f_{\lambda} ( \bk_2 , t)][ 1 + f_{\lambda} ( \bk_3 , t)] \nonumber \\
&&~~~~~~~~~~~~+  (n-2) f_{\lambda} ( \bk , t)  n (\varepsilon_{k_1} )
[ 1 + f_{\lambda} ( \bk_2 , t)][ 1 + n (\varepsilon_{k_3} )] \nonumber \\
&&~~~~~~~~~~~~+ \frac{(n-2)}{2} f_{\lambda} ( \bk , t) f_{-\lambda} (\bk_1 , t) 
[ 1 + n (\varepsilon_{k_2} )][ 1 + n (\varepsilon_{k_3} )] \nonumber \\
&&~~~~~~~~~~~~- 4 [1 + f_{\lambda} ( \bk , t)][1 + f_{-\lambda} (\bk_1 , t)]
f_{\lambda} ( \bk_2 , t) f_{-\lambda} ( \bk_3 , t) \nonumber \\
&&~~~~~~~~~~~~- 2 [1 + f_{\lambda} ( \bk , t)][1 + f_{\lambda} (\bk_1 , t)]
f_{\lambda} ( \bk_2 , t) f_{\lambda} ( \bk_3 , t) \nonumber \\
&&~~~~~~~~~~~~-  (n-2) [1 + f_{\lambda} ( \bk , t)][1 + n (\varepsilon_{k_1} )]
f_{\lambda} ( \bk_2 , t) n (\varepsilon_{k_3} ) \nonumber \\
&&~~~~~~~~~~~~-  \frac{(n-2)}{2} [1 + f_{\lambda} ( \bk , t)][1 + f_{-\lambda} (\bk_1 , t)]
n (\varepsilon_{k_2} ) n (\varepsilon_{k_3} )
\Bigr\}
\end{eqnarray}
with $\lambda = \pm 1$.
The analysis of the linearized form of this equations is similar to that for (\ref{trans1}),
but will not be presented here.
Such an analysis will lead to a determination of the spin conductivity, $\sigma$,
which is related to the spin diffusion constant $D_s$ by the Einstein relation
\begin{equation}
\sigma = D_s \chi ,
\end{equation}
where $\chi$ is the uniform spin susceptibility. Results for the $\chi$ at non-zero
$T$ above the critical point have been given earlier in the $1/n$~\cite{cs,csy}
and $\epsilon$~\cite{s} expansions.

\section{Computations with the collision term}
\label{sec:collision}

We describe here some of 
the steps in between the original transport equation (\ref{trans1})
and the linearized form (\ref{inteq}).

First we insert the parametrization (\ref{cless2}) into (\ref{trans1}), and linearize
in the the external electric field. Then, notice that in the collision term the unknown
function $\psi$ appears only in the integrals over the radial components of the momenta. 
The angular integrals involve only known functions and can be performed analytically.
As already noted in Section~\ref{sec:ctrans}, the integrals in the collision
term can be done directly in $d=3$ to obtain the leading result, and so all
the computations here are for $d=3$. One needed angular integral is
\begin{eqnarray}
\int dk_3 g(k_3) \int d \Omega_{\bk_1} d \Omega_{\bk_2} d \Omega_{\bk_3} 
&&\delta^3 ( \bk + \bk_3 - \bk_1 - \bk_2 ) \delta(k + k_3 - k_1 - k_2) \nonumber \\
&&~~~~~~~~ = 
\frac{8 \pi^2}{k k_1 k_2 (k_1 + k_2 - k)} g(k_1+k_2 -k) I_1 ( k, k_1, k_2),
\label{ac1}
\end{eqnarray}
where $g(k_3)$ is some function of $k_3$. The result involves the function
$I_1$ which is given by 
\begin{equation}
I_1(k, k_1, k_2) = \left\{ \begin{array}{ccl}
0 & & \mbox{$k_1 + k_2 \leq k$} \\
k_1 + k_2 - k  & &\mbox{$k_1 + k_2 \geq k$ and $k_1 \leq k$ and $k_2 \leq k$} \\
k_1 & ~{\rm for}~ &\mbox{$k_1 \leq k$ and $k_2 \geq k$} \\
k & &\mbox{$k_1 \geq k$ and $k_2 \geq k$} \\
k_2 & &\mbox{$k_1 \geq k$ and $k_2 \leq k$}
\end{array} . \right.
\end{equation}
A second angular integral has a single vector momentum in the integrand
\begin{eqnarray}
\int dk_3 g(k_3) \int d \Omega_{\bk_1} d \Omega_{\bk_2} d \Omega_{\bk_3} 
\bk_1 && \delta^3 ( \bk + \bk_1 - \bk_2 - \bk_3 ) \delta(k + k_1 - k_2 - k_3) \nonumber \\
&&~~~~ = -
\frac{8 \pi^2 \bk}{3 k^3 k_1 k_2 (k + k_1 - k_2)} g(k+k_1 -k_2) I_2 ( k, k_1, k_2),
\label{ac2}
\end{eqnarray}
where  now
\begin{equation}
I_2 (k, k_1, k_2) = \left\{ \begin{array}{ccl}
0 & & \mbox{$k_2 \geq k + k_1$} \\
k^3 + k_1^3 + 2k_2^3 - 3 k_2^2 ( k + k_1) + 3 k k_1 k_2
  & &\mbox{$k_2 \leq k + k_1$ and $k_2 \geq k$ and $k_2 \geq k_1$} \\
k^3 & ~{\rm for}~ &\mbox{$k_2 \geq k$ and $k_2 \leq k_1$} \\
-2k_2^3 + 3 k_2^2 ( k + k_1) - 3 k k_1 k_2 & &\mbox{$k_2 \leq k$ and $k_2 \leq k_1$} \\
k_1^3 & &\mbox{$k_2 \geq k_1$ and $k_2 \leq k$}
\end{array}. \right.
\end{equation}
Finally, we also need a variation of (\ref{ac2}): 
\begin{eqnarray}
\int dk_3 g(k_3) \int d \Omega_{\bk_1} d \Omega_{\bk_2} d \Omega_{\bk_3} 
\bk_1 && \delta^3 ( \bk + \bk_2 - \bk_1 - \bk_3 ) \delta(k + k_2 - k_1 - k_3) \nonumber \\
&&~~~~ = 
\frac{8 \pi^2 \bk}{3 k^3 k_1 k_2 (k + k_2 - k_1)} g(k+k_2 -k_1) I_3 ( k, k_1, k_2), 
\label{ac3}
\end{eqnarray}
where  
\begin{equation}
I_3 (k, k_1, k_2) = \left\{ \begin{array}{ccl}
0 & & \mbox{$k_1 \geq k + k_2$} \\
k^3 - k_1^3 - 2k_2^3 - 3 k_2^2 ( k - k_1) + 3 k k_1 k_2
  & &\mbox{$k_1 \leq k + k_2$ and $k_1 \geq k$ and $k_2 \leq k_1$} \\
k^3 & ~{\rm for}~ &\mbox{$k_1 \geq k$ and $k_2 \geq k_1$} \\
-2k_2^3 - 3 k_2^2 ( k - k_1) + 3 k k_1 k_2 & &\mbox{$k_1 \leq k$ and $k_2 \leq k_1$} \\
k_1^3 & &\mbox{$k_2 \geq k_1$ and $k_1 \leq k$}
\end{array} . \right.
\end{equation}

Using (\ref{ac1}-\ref{ac3}) in the linearized form of (\ref{trans1}), the integral
equation satisfied by $\psi$ becomes
\begin{eqnarray}
&& - i \omega \psi(k, \omega ) + \frac{1}{k} \frac{\partial n ( k )}{
\partial k}  \nonumber \\
&&~~~= - \frac{\pi \epsilon^2}{75 k^4} \left\{
\frac{18 \psi(k,\omega) k^2}{n(k)} \int_0^{\infty} dk_1 dk_2
I_1 ( k, k_1, k_2 ) n(k_2) n(k_1) [1 + n(k_1 + k_2 - k)]
\right. \nonumber \\
&&~~~~~~~~~~~+2[1 + n(k)] \int_0^{\infty} dk_1 dk_2 \frac{\psi(k_1 , \omega)}{n(k_1)}
I_2 (k, k_1 , k_2) n(k_2) n(k+ k_1 - k_2 ) \nonumber \\
&&~~~~~~~~~~~\left. -4 n(k) \int_0^{\infty} dk_1 dk_2 \frac{\psi(k_1 , \omega)}{n(k_1)}
I_3 (k, k_1 , k_2) n(k_2)[1+ n(k+ k_2 - k_1 )] \right\}.
\label{appcoll}
\end{eqnarray}

This can be turned into the form (\ref{inteq}) by evaluating the integrals
over $k_2$, which can also be done analytically. To do this, it is first necessary
to separate the products of two Bose functions each with $k_2$ in their arguments,
into terms in which only one Bose function involves $k_2$: this is done by
repeated use of the identity
\begin{equation}
n(k_2 + a) n(k_2 + b) = n(b-a)n(k_2 + a) + n(a-b) n(k_2 + b)
\end{equation}
which is valid for any $a$, $b$.
From the forms of the functions $I_{1,2,3}$ above, it is now clear that we only need integrals
of $k_2$ over Bose functions of $k_2$, times powers of $k_2$. The general integral
needed is of the form
\begin{equation}
\int_a^b d k_2 \frac{ k_2^n}{e^{(k_2 + c)/T} - 1} = 
\int_a^{\infty} d k_2 \frac{ k_2^n}{e^{(k_2 + c)/T} - 1}
-\int_b^{\infty} d k_2 \frac{ k_2^n}{e^{(k_2 + c)/T} - 1}
\end{equation}
where $n$ is an integer, and $c$ is a real number. Changing variables on the right hand
side, we can rewrite the first term on the right hand side as
\begin{equation}
\int_a^{\infty} d k_2 \frac{ k_2^n}{e^{(k_2 + c)/T} - 1} = 
\int_0^{\infty} d k_2 \frac{ (k_2+a)^n}{e^{(k_2 + c + a)/T} - 1},
\end{equation}
and similarly for the second term. Expanding the polynomial in the numerator, we finally
conclude that all of the integrals over $k_2$ can be reduced to the following basic
integral:
\begin{equation}
\int_0^{\infty} dk_2 \frac{k_2^n}{e^{(k_2 + a)/T} - 1} = T^{n+1} \Gamma(n+1) {\rm Li}_{n+1}
(e^{-a/T}).
\end{equation}
Here ${\rm Li}_p (z)$ is the polylogarithm function of order $p$, defined
by the series
\begin{equation}
{\rm Li}_p (z) = \sum_{n=1}^{\infty} \frac{z^n}{n^p}.
\end{equation}
Notice ${\rm Li}_p ( 1) = \zeta ( p )$.

After performing the integrals over $k_2$ as described above, (\ref{appcoll}) takes
the form (\ref{inteq}). We will not display explicit expressions for $F_{1,2}  (k, k_1)$
here as they are quite lengthy and not particularly informative.
We used a CERNLIB routine for numerical evaluation
of ${\rm Li}_p ( z )$ for $0 < z < 1$: this allowed very rapid determination
of the kernel of the integral equation (\ref{inteq}).

\section{Order parameter susceptibility}
\label{susc}

This appendix will review the results of Ref.~\onlinecite{s} on the finite temperature
crossovers in the order parameter susceptibility $\chi = \langle \phi_{\alpha}
\phi_{\alpha} \rangle$ (no summation over $\alpha$). Ref.~\onlinecite{s} was
concerned with making statements correct to all orders in $\epsilon$ and to understanding
the crossover to classical critical fluctuations in the vicinity of the finite temperature
phase transition line, and this required a rather involved analysis. Here, we are mainly 
interested in low order results at 
finite temperatures
above the quantum-critical point and the quantum disordered phase: in this case the
necessary results can be obtained directly in a self-consistent Hartree-Fock-like analysis, 
as we now show.

The self-consistent Hartree-Fock susceptibility of ${\cal S}$ is obtained by summing
all the one-loop tadpole diagrams; this leads to the expression
\begin{equation}
\chi (k, \omega_n) = \frac{1}{\omega_n^2 + k^2 + m^2 (T)}
\end{equation}
where $m^2 (T)$ is given by the solution of the equation
\begin{eqnarray}
m^2 (T) &=& t_0 + m_{0c}^2 + u_0 \left( \frac{n+2}{6} \right)
T \sum_{\epsilon_n} \int \frac{d^d q}{(2 \pi)^d} \frac{1}{\epsilon_n^2
+ q^2 + m^2 (T)} \nonumber \\
&=& t_0 + u_0 \left( \frac{n+2}{6} \right) \left[
T \sum_{\epsilon_n} \int \frac{d^d q}{(2 \pi)^d} \frac{1}{\epsilon_n^2
+ q^2 + m^2 (T)} - \int \frac{d^{d+1}p}{(2 \pi)^{d+1}} \frac{1}{p^2} \right]
\label{a1}
\end{eqnarray}
In the second equation we have inserted the leading result for the value of 
$m_{0c}^2$. We are following the convention here of denoting $d+1$ dimensional
spacetime momenta by $p$, and $d$ dimensional spatial momenta by $k,q$. 
In the critical region, the coupling $u_0$ is of order $\epsilon$, so one might think
that it is permissible to set $m^2 (T) = t_0$ on the right hand side of (\ref{a1}) and
obtain a result correct to order $\epsilon^2$. However this is not adequate for
our purposes for two reasons~\cite{s}: \\
({\em i}) The resulting expression for
$m^2 (T)$ is not analytic as a function of $t_0$ at $t_0 = 0$ for $T> 0$. This analyticity
is required on the physical ground that there can be no thermodynamic singularity
at $T>0$ at the quantum-critical coupling $t_0 = 0$.\\
({\em ii}) At $T>0$ above the quantum critical coupling $t_0$ there is a contribution
to $m^2 (T)$ of order $\epsilon^{3/2}$ which is missed.

To proceed, we can either use the analysis of Ref~\onlinecite{s}, or directly analyze
the singularity structure of (\ref{a1}) as $u_0 \rightarrow 0$. By either method,
it can be shown that it is permissible to set $m^2 (T) = t_0$ in only the
$\epsilon_n \neq 0$ terms on the right hand side of (\ref{a1}). To describe this, we
write
\begin{equation}
m^2 (T) = R(T) + \delta m^2 (T)
\label{a2}
\end{equation}
where
\begin{equation}
R(T) = t_0 + u_0 \left( \frac{n+2}{6} \right) \left[
 \int \frac{d^d q}{(2 \pi)^d} \left( T \sum_{\epsilon_n \neq 0} \frac{1}{\epsilon_n^2
+ q^2 + t_0} + \frac{T}{q^2} \right)
- \int \frac{d^{d+1}p}{(2 \pi)^{d+1}} \frac{1}{p^2} \right]
\label{a3}
\end{equation}
was a quantity introduced in Ref~\onlinecite{s}, and 
\begin{equation}
\delta m^2 (T) = u_0 \left( \frac{n+2}{6} \right) 
T \int \frac{d^d q}{(2 \pi)^d} \left( \frac{1}{
q^2 + m^2 (T)} - \frac{1}{q^2} \right).
\label{a4}
\end{equation}
Finally, on the right hand side of (\ref{a4}) we replace $m^2 (T)$ by $R(T)$.
The resulting expression for $\delta m^2 (T)$ when inserted along with (\ref{a3}) into
(\ref{a2}) gives us our final result for $m^2 (T)$.

We will now manipulate the above result for $m^2 (T)$ into a form which
explicitly displays its universal scaling nature. First replace the bare
couplings by renormalized couplings; to the order we are computing things,
this is equivalent to performing the substitutions~\cite{bgz,zj} 
$u_0 = \mu^{\epsilon} g/S_{d+1}$ and $t_0 = t( 1 + (n+2) g/(6 \epsilon))$ where
$\mu$ is a renormalization momentum scale, $g$ is a dimensionless coupling constant, and $t$ is 
a renormalized coupling measuring the deviation from the critical point. Expanding the
result to order $g$, we find that the poles in $\epsilon$ cancel. Finally, we set
$g$ at its fixed point value~\cite{bgz,zj} $ g = g^{\ast} = 6 \epsilon / (n+8)$.
Our final result for $m^2 (T)$ then becomes
\begin{equation}
m^2 (T) = R(T) - \epsilon \left( \frac{n+2}{n+8} \right) 2 \pi T \sqrt{R(T)}
\label{a4a}
\end{equation}
where the second term on the right hand side is the contribution of $\delta m^2 (T)$.
A somewhat subtle analysis, discussed at some length in Ref~\onlinecite{s},
is required to obtain the following final form for $R(T)$:
\begin{equation}
R(T) = t \left( 1 + \epsilon \frac{n+2}{n+8} \ln \frac{T}{\mu} \right)
+ \epsilon T^2 \frac{n+2}{n+8} G \left( \frac{t}{T^2} \right),
\label{a5}
\end{equation}
with the crossover function $G(y)$ given by
\begin{equation}
G(y) = - 2 \int_0^{\infty} dq \left[
\ln \left( 2 q^2 \frac{(\cosh (\sqrt{q^2 + y}) - 1)}{q^2 + y} \right)
-q - \frac{y}{2 \sqrt{q^2 + 1/e}} \right]
\end{equation}
This form of $G(y)$ is valid for both negative and positive $y$ (when the argument
of the square root is negative we use the identity $\cosh ix = \cos x$)
and is easily shown to be analytic at $y=0$ where
\begin{equation}
G(0) = \frac{2 \pi^2}{3}~~~~,~~~~dG/dy (0) = 2.453808582\ldots .
\end{equation}
The result (\ref{a5}) can therefore be used on both sides of the critical coupling
$t=0$, and the required analyticity of $T>0$ properties at the critical coupling
has been achieved. Also notice that at $t=0$, $R(T)$ is of order $\epsilon$,
so the result (\ref{a4a}) contains a term of order $\epsilon^{3/2}$.

Finally, let us express $m^2 (T)$ in terms of experimentally measurable energy
scales. The energy scales have to be defined differently for $t>0$ and $t<0$.

For $t>0$ we choose the $T=0$ energy gap, $\Delta$, to measure the deviation from the
critical point. Using the relation~\cite{s} $\Delta^2 = \mu^2 ( t/\mu^2 )^{2 \nu}$
in (\ref{a5}) ($\nu = 1/2 + \epsilon (n+2)/(4 (n+8))$ is the correlation length
exponent) we get our final universal expression for $R(T)$, valid everywhere above
the quantum-disordered (insulating) phase ($t>0$)
\begin{equation}
R(T) = \Delta^2 \left( 1 + \epsilon \frac{n+2}{n+8} \ln \frac{T}{\Delta} \right)
+ \epsilon T^2 \frac{n+2}{n+8} G \left( \frac{\Delta^2}{T^2} \right)
\end{equation}
Notice that the arbitrary momentum scale $\mu$ has disappeared, and combined 
with (\ref{a4a}) we now have a result for $m^2 (T)$ expressed solely in terms
of the measurable energy scales $\Delta$ and $T$.

For $t<0$, our results are confined to the normal phase $T > T_c (t)$.
We measure the deviation from the critical point by the value of the
superfluid stiffness, $\rho_s$ at $T=0$ (for $n=3$ $\rho_s$ is the spin stiffness
of the ordered antiferromagnetic phase). To obtain a quantity with the
dimensions of energy we define
\begin{equation}
\widetilde{\rho}_s \equiv \left((3-d) \rho_s\right)^{1/(d-1)}.
\end{equation}
The numerical factors are for future convenience; also notice that
in $d=2$, $\widetilde{\rho_s} \equiv \rho_s$.
Then we use the expression for $\rho_s$ in terms of $t$ and $\mu$ in
Ref~\onlinecite{s} to express $t$ in terms of $\widetilde{\rho}_s$ and $\mu$
in (\ref{a5}). Then all the $\mu$ dependencies cancel as before, 
and we get our final result
\begin{eqnarray}
R(T) = - \frac{\widetilde{\rho}_s^2}{n+8} && \left(
1 - \frac{\epsilon}{2(n+8)} + \frac{3 \epsilon}{n+8} \ln \frac{2}{n+8}
+ \epsilon \frac{n+2}{n+8}
 \ln \frac{T}{\widetilde{\rho}_s} \right) \nonumber \\
&&~~~~~~~~~~~~
+ \epsilon T^2 \frac{n+2}{n+8} G \left( -\frac{\widetilde{\rho}_s^2}{(n+8) T^2} \right)
\end{eqnarray}
Again combined with (\ref{a4a}) we now have an expression for $m^2 (T)$ in terms
of the superfluid stiffness of the ordered phase at $T=0$ for $t<0$
and $T > T_c (t)$.

\begin{references}
\bibitem{luttinger}
It is worth noting here a key difference between our analysis of
${\cal S}$ in $d=2$,
and the non-zero temperature transport of $d=1$
Luttinger liquids as discussed 
in Ref~\protect\cite{gm}. The bosonic formulation of the latter
model has a boson with a single flavor; the momentum and group velocity
are not equal, and so the total momentum and charge current are not
proportional to each other for an arbitrary distribution function of the particles. 
Nevertheless, in
this Luttinger model,  there is zero resistance
in the absence of umklapp scattering~\cite{gm}. 
The crucial distinction is that ${\cal S}$ has $n>1$ flavors
of bosons (see Section~\protect\ref{qtrans}) 
while the model of Ref~\protect\cite{gm} has only one.
As a result, any current carrying state of the latter model has 
a non-zero momentum which cannot decay, while it is possible for the former
model to have a current carrying state with vanishing momentum.
So the important point is not just that the total momentum and charge
current of ${\cal S}$ are not the same, but that this difference arises
from the presence of $n>1$ flavors of bosons.
\end{references}
\end{document}